\documentclass{aa}  

\usepackage{graphicx}
\usepackage{txfonts}
\begin{document}

   \title{Earliest phases of star formation (EPoS):}

   \subtitle{Dust temperature distributions in isolated starless cores}

   \author{N. Lippok\inst{1}\and
	   R. Launhardt\inst{1}\and
	   Th. Henning\inst{1}\and
	   Z. Balog\inst{1}\and
	   H. Beuther\inst{1}\and
	   J. Kainulainen\inst{1}\and
	   O. Krause\inst{1}\and
	   H. Linz\inst{1}\and
	   M. Nielbock\inst{1}\and
	   S. E. Ragan\inst{1,2}\and
	   T. P. Robitaille\inst{1}\and
	   S. I. Sadavoy\inst{1}\and
	   A. Schmiedeke\inst{1,3}
          }

   \institute{Max-Planck-Institut f\"ur Astronomie (MPIA), K\"onigstuhl 17, D-69117 Heidelberg, Germany\\ \email{rl@mpia.de}\label{inst1}
   \and University of Leeds, Leeds, LS2 9JT, UK\label{inst2}
    \and Universit\"at zu K\"oln, Z\"ulpicher Str. 77, D-50937 K\"oln, Germany\label{inst3}
         }

   \date{Received; Accepted}

\abstract{Constraining the temperature and density structure of dense molecular cloud cores is fundamental for understanding 
the initial conditions of star formation. 
We use \textit{Herschel} observations of the thermal FIR dust emission from nearby and isolated 
molecular cloud cores and combine them with ground-based submillimeter continuum data to derive observational 
constraints on their temperature and density structure.}
{The aim of this study is to verify the validity of a ray-tracing inversion technique developed to derive the 
dust temperature and density structure of nearby and isolated starless cores directly from the dust emission maps 
and to test if the resulting temperature and density profiles are consistent with physical models.}
{Using this ray-tracing inversion technique, we derive the dust temperature 
and density structure of six isolated starless cloud cores. 
We employ self-consistent radiative transfer modeling to the derived density profiles,  
treating the ISRF as the only heating source. 
The best-fit values of local strength of the ISRF and the extinction by the outer envelope are derived by comparing 
the self-consistently calculated temperature profiles with those derived by the ray-tracing method.}
{We find that all starless cores are significantly colder inside than outside, 
with the core temperatures showing a strong negative correlation with peak column density.
This suggests that 
their thermal structure is dominated by external heating from the ISRF and shielding by dusty envelopes.
The temperature profiles derived with the ray-tracing inversion method can be 
well-reproduced with self-consistent radiative transfer models.
We confirm results from earlier studies that found that the usually 
adopted canonical value of the total strength of the ISRF in the solar 
neighbourhood is incompatible with the most widely used dust opacity models 
for dense cores. However, with the data available, we cannot 
uniquely resolve the degeneracy between dust opacity law and strength of the ISRF.}
{}


   \maketitle
%

\section{Introduction}   \label{sec_intro}

Stars ultimately form through the gravitational collapse of cold and dense molecular cloud cores, 
irrespective of how these cores were formed in the first place.
The initial temperature and density structure of such gravitationally 
bound cloud cores are important properties that determine the onset and course of the collapse. 
Yet, until very recently we had little direct observational evidence of the internal temperature structure of such cores.
Consequently, the derivation of density profiles from dust emission data was often done with the simplifying 
assumption of isothermality \citep[e.g.,][]{ward-thompson1999,kirk2005,launhardt2005,launhardt2010}
or relied on weakly constrained model temperature profiles \citep[e.g.,][]{evans2001,zucconi2001,shirley2005}.
Only more recently have some studies tried to constrain the temperature distribution of starless molecular 
cloud cores directly from observational data of the thermal dust continuum emission 
\citep[e.g.,][]{ward-thompson2002,schnee2005,stamatellos2007,stutz2010,nielbock2012,beuther2012,lippok2013,launhardt2013,pitann2013,schmalzl2014}.

The thermal emission from dust grains is indeed the most robust tracer of the temperature and density structure 
of such cold and dense cloud cores \citep[see discussion in][]{launhardt2013}. 
At the typical temperature of starless cores, 6\,--\,20\,K, the dust grains emit thermal radiation mainly at 
far-infrared (FIR) wavelengths. The spectral shape of this emission depends on the temperature (and density) 
distribution along the line-of-sight (LOS), the optical depth of the emitting layer, and the properties of the dust grains. 
Longward of \mbox{$\lambda\approx200\,\mu$m}, the thermal emission is usually optically thin even 
at column densities of $10^{25}$\,cm$^{-2}$\ and thus traces well the structure in the interior of the dense cores,
provided the dust temperature and the optical properties of the grains are known.

Thus, to derive the density structure from the thermal dust emission,
cores must be observed in the FIR, in particular, toward the peak of their thermal spectral energy distribution (SED).  
Since the FIR is difficult to observe 
from the ground, most previous studies generally lacked these key observations or had to rely on relatively low-resolution 
FIR flux measurements from space observatories like 
{\it IRAS} (1\arcmin--2\arcmin\ at 60--100\,$\mu$m), 
{\it ISO} (1\arcmin--3\arcmin\ at 60--200\,$\mu$m), 
{\it Spitzer} (18\arcsec--40\arcsec\ at 70--160\,$\mu$m), 
or {\it AKARI} (60\arcsec--90\arcsec\ at 65--160\,$\mu$m).
The \textit{Herschel}\footnote{\textit{Herschel} is an ESA space observatory with science instruments provided by 
European-led Principal Investigator consortia and with important participation from NASA.} 
space observatory \citep{pilbratt2010} was the first facility to cover most of the FIR wavelength range with 
high sensitivity and at an angular resolution \citep[6\arcsec--36\arcsec\ at 70--500\,$\mu$m;][]{aniano2011}
that is comparable to the largest ground-based single-dish millimeter telescopes.
Thus, with \textit{Herschel} observations and complementary long-wavelengths ground-based data, 
we are able to derive the dust temperature structure of molecular cloud cores and to put more robust constraints 
on their density structure than what was possible in the pre-\textit{Herschel} era.

With the goal of constraining the physical conditions in molecular cloud cores during the prestellar and the earliest 
protostellar stages of star formation, we initiated the \textit{Herschel} guaranteed time key project "earliest phases 
of star formation'' \citep[EPoS;][]{ragan2012,launhardt2013}. As part of this project, we observed a sample of Bok globules at five continuum wavelengths between 
100\,$\mu$m and 500\,$\mu$m. Bok globules are small, nearby, and relatively isolated molecular clouds with a single core 
or at most very few dense cores, which makes them 
the most simply-structured and easily observable star-forming units in the Galaxy and ideal laboratories to study in detail 
the initial conditions of isolated star formation. 
An important selection criterion for the globules targeted in the EPoS 
survey was that they are located in regions with exceptionally low FIR background confusion noise to allow for deep 
observations at 100\,$\mu$m. This wavelength is crucial for deriving the temperature of the 
cold dust since it constrains the short-wavelength side of the SED peak. This also implies that our targets are all 
located outside the Galactic plane at latitudes of $|b|\approx3.4\ldots12.4\degr$.

Using a robust temperature mapping algorithm that applies modified black-body fits to the dust emission spectra, we have 
already shown that these isolated cores are thermally dominated by heating from the interstellar radiation field (ISRF); 
they have warm (14\,--\,20\,K) exteriors and cooler interiors \citep[$<$\,11\,--\,14\,K,][]{stutz2010,launhardt2013}. 
In addition, because of their isolation and simple structure, we were able to use a ray-tracing inversion technique to reconstruct the 
3-D temperature and density profiles of the globules \citep{nielbock2012,lippok2013,schmalzl2014}.  
We found that the temperature profiles of the starless cores drop to 7\,--\,13\,K  toward their centers, 
while their outer detectable rims are typically warmer by 5.5$\pm$2.5\,K.

In this paper, we compare the dust temperature profiles derived using this ray-tracing inversion technique with self-consistent radiative 
transfer models and explore the physical conditions that lead to the observed temperature distributions.
The paper is structured as follows. 
In Section\,\ref{sec_data}, we describe the sources and data used in this study. 
In Section~\ref{sec_mod}, we describe the overall strategy as well as the dust model and modeling tools. 
In Section\,\ref{sec_res}, we compare the results for the dust temperature profiles derived by 
the ray-tracing inversion method with those predicted by the self-consistent radiative transfer models. 
The results and uncertainties are discussed in Section\,\ref{sec_disc}.
Finally, Section\,\ref{sec_sum} summarizes and concludes the paper.


\section{Sources and data}     \label{sec_data}

\begin{table*}[htb]
\small
\caption{\label{t1}Source list}
\centering
\begin{tabular}{llcllcll}
\hline\hline
Source & Other   & R.A., Dec. (J2000)\tablefootmark{a}  & Energetic                         & Source                                & Dist. & Region & Ref.\\
              & names &[h:m:s, $^\circ$:$\arcmin$:$\arcsec$] & status\tablefootmark{b} & notes\tablefootmark{c}    &  [pc]  &               &  \\
\hline
CB\,4\,-\,SMM                                  &  $\ldots$         & 00:39:05.2, +52:51:47   & sub-critical    & 1     & $350\pm 150$ &  Cas\,A, Gould's Belt (GB) & 1,2\\
CB\,17\,-\,SMM\tablefootmark{c}     & L\,1389           & 04:04:37.1, +56:56:02    & $\sim$stable & 2, 3 & $250\pm 50$  &  Perseus, GB                        & 2,3,1\\
CB\,26\,-\,SMM2\tablefootmark{d}   & L\,1439            & 05:00:14.5, +52:05:59   & $\sim$stable & 2, 4 & $140\pm 20$  &  Auriga                                  & 2,3,4,5\\
CB\,27\,-\,SMM                                & L\,1512-S        & 05:04:08.1, +32:43:30    & $\sim$stable & 1 & $140\pm 20$  &  Taurus-Auriga                    & 2,4,5\\
B\,68\,-\,SMM                                   & L\,57, CB\,82 & 17:22:38.3, $-$23:49:51 & $\sim$stable & 1 & $135\pm15$  &  Ophiuchus, Pipe nebula  & 2,6,7,8 \\
CB\,244\,-\,SMM2\tablefootmark{e} & L\,1262           & 23:25:26.8, +74:18:22    & super-critical & 2, 5 & $200\pm30$  &  Cepheus flare, GB             & 2,3,4,9 \\
\hline
\end{tabular}
\tablefoot{
\tablefoottext{a}{Position of the column density peak of the starless core as derived in \citet{launhardt2013}.} 
\tablefoottext{b}{Based on the stability analysis in Sect.\,\ref{ssec_res_rti} (Fig.\,\ref{fig_massdens}).}
\tablefoottext{c}{Source notes: 1: single-core globule, 2: double-core globule,
                            3: CB\,17 contains an additional low-luminosity Class\,I YSO (IRS) located $25\arcsec$\ from starless core (SMM), not embedded, but partially blending,
                            4: CB\,26 contains an additional Class\,I YSO (SMM1) located $3.6\arcmin$\ south-west of the starless core (SMM2),
                            5: CB\,244 contains an additional Class\,0 source (SMM1) located $\sim 90\arcsec$\ east of starless core (SMM2), partially blending.}
}
\tablebib{(1)~\citet{perrot2003}; (2)~\citet{launhardt2013}; (3)~\citet{launhardt2010}; (4)~\citet{loinard2011}; 
  (5)~\citet{stutz2009}; (6)~\citet{deGeus1989};
  (7)~\citet{lombardi2006}; (8)~\citet{alves2007}; (9)~\citet{kun1998}.
}
\end{table*}

Based on the results of earlier studies, we selected for the EPoS survey 
12 nearby, isolated, and previously well-characterized Bok globules, all located in regions 
of exceptionally low cirrus confusion noise \citep[see][and references therein]{launhardt2013}.
Each of these globules contains one to two embedded cores which, in some cases, 
are of different nature or evolutionary status.
For this study of the temperature structure of starless cores, we selected those six globules from the 
EPoS sample that contain at least one starless core.
We note that three of the six selected globules also contain a protostar in addition to the starless core
(see Tab.\,\ref{t1} and Fig.\,\ref{fig-tmaps}). However, for this study we mask out the protostars (see Sect.\,\ref{ssec_res_rti}),
analyze the starless cores only, and discuss the uncertainties introduced by the presence of the nearby protostars 
in Sect.\,\ref{ssec_disc_uncert}. Throughout this paper, we use the same core name convention as in \citet{launhardt2013},
except where explicitly referring to the entire globule or where distinction between core and globule in single-core globules 
would be meaningless.
Source names, coordinates, and distances of the selected cores are summarized in Table\,\ref{t1}.
Total gas masses and outer radii of the cores are in the ranges \mbox{2.6\,--\,14\,M$_{\odot}$}\ 
and \mbox{0.2\,--\,0.4\,pc} (see Table\,\ref{t_coreproperties}). 
We use maps of the dust continuum emission in the PACS 100\,$\mu$m, 160\,$\mu$m \citep{poglitsch2010}, 
and SPIRE 250\,$\mu$m, 350\,$\mu$m, and 500\,$\mu$m bands \citep{griffin2010}
that were obtained as part of the \textit{Herschel} guaranteed time key project EPoS 
and processed as described in \citet{launhardt2013}. These data are complemented by ground-based 
(sub-)mm observations at 450\,$\mu$m, 850\,$\mu$m, and 1.2\,mm that are presented in \citet{launhardt2010,launhardt2013} 
and references therein.


\section{Modeling methods}    \label{sec_mod}


\subsection{General strategy}    \label{ssec_mod_strat}

The ray-tracing inversion technique was developed to infer the dust temperature and density 
structure of starless cores directly from the observed dust emission maps without the need to make 
assumptions about the specific physical conditions \citep{nielbock2012}. 
Apart from scattering (which is neglected by the ray-tracing), 
both the ray-tracing inversion and self-consistent radiative transfer 
use the same basic equations to relate the dust properties (opacity, temperature, and optical depth) to the effectively 
emitted flux density.
However, the two methods have different approaches: The ray-tracing 
inversion assumes pre-defined analytical parameterized prescriptions for the temperature and density profiles 
before iteratively optimizing their parameter values by independently fitting the flux densities (SEDs) at each map pixel. 
The profile equations and iteration method are described in Sect.\,\ref{ssec_mod_rti}
\citep[see also][]{nielbock2012,lippok2013,schmalzl2014}.
The self-consistent radiative transfer, on the other hand, also uses pre-defined density profiles (e.g., from a physical model or here derived by the ray-tracing 
inversion technique), but involves a model of the ISRF to calculate self-consistently the corresponding equilibrium temperature distribution.

The ray-tracing technique has the advantage that it does not need to make assumptions about the physical conditions,
but only assumes a dust opacity model and free parameterized profile shapes for the temperature and density.
It can also capture deviations from spherical symmetry and reproduce more complicated source structures.
With this approach, a large and densely sampled parameter space can be probed with relatively little computational effort.
On the other hand, the ray-tracing technique does not directly provide constraints on the physical conditions (e.g., the ISRF) 
that lead to the observed temperature and density distributions. 
This short-coming can, however,  also be advantageous when compared to self-consistent radiative transfer modeling.
If the range of parameters probed by a given setup of forward modeling is too small, 
the resulting 'best-fit' model would not reveal the actual physical conditions that characterize the observed target. 
In contrast, the ray-tracing inversion technique yields temperature and density profiles 
that are likely closer to the real distributions and may thus lead to better hints toward the actual physical conditions
when afterwards verified by self-consistent radiative transfer models.
Thus, the ray-tracing inversion technique is well-suited to exploring the dust emission data and derive the source structures, 
but should then be complemented by a radiative transfer model to better characterize  
the physical conditions and environments of the cores.

For exploring our {\it Herschel} and complementary ground-based continuum data of isolated molecular cloud cores,
we therefore chose to first apply the ray-tracing inversion technique, and afterwards verify the results by 
self-consistent radiative transfer models. 
As we show in Sect.\,\ref{ssec_mod_rti}, the analytical parameterization of the radial density profiles 
used for the ray-tracing inversion indeed reflects the actual physical configurations of the cores and the solutions found 
by the ray-tracing inversion for both the density and the temperature is practically always unique. 
With the results in hand, we now determine the physical conditions that best match the inferred temperature profiles. 
For this purpose, we adopt the azimuthally averaged density profiles derived with the ray-tracing inversion as input for 
self-consistent 1-D radiative transfer modeling and calculate the equilibrium temperature distributions for 
these density profiles. We use the same dust opacity model as in the ray-tracing inversion 
(Sect.\,\ref{ssec_mod_dust}), but vary both the total strength of the ISRF 
and the extinction by a surrounding envelope, as described in Sect.\,\ref{ssec_mod_srt}.
We then determine how well and for what combination of ISRF strength and envelope extinction 
the predicted equilibrium temperature profiles agree with the ray-tracing results.


\subsection{The dust model}    \label{ssec_mod_dust}

\begin{figure}[htb]
   \centering
   \includegraphics[width=0.45\textwidth]{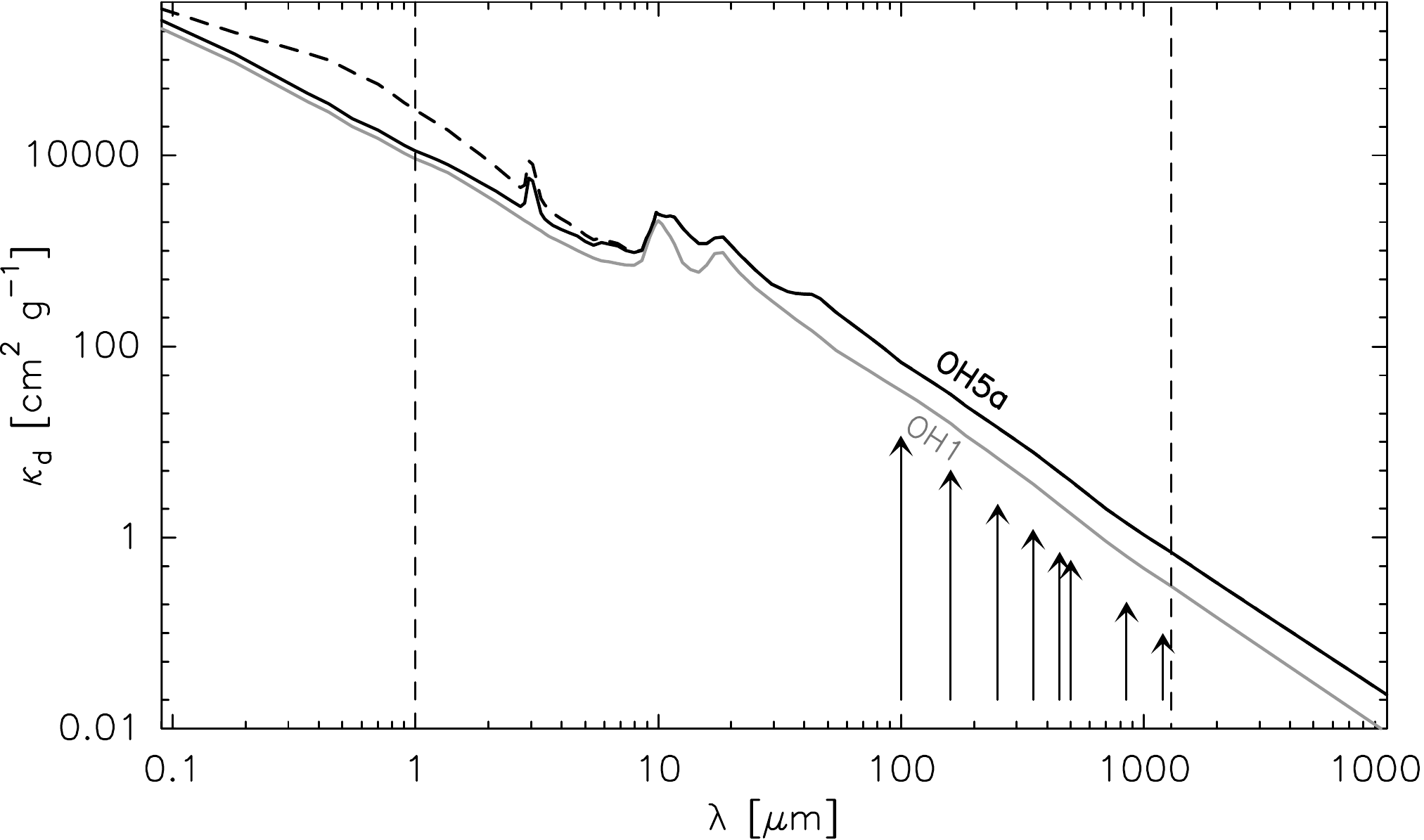}
   \caption{Dust opacity models (extinction/absorption coefficient per g of dust) used in this paper.
   Black lines show the absorption (solid) and extinction (dashed) coefficients of the modified 
   OH5a model \citep{oh94}. The gray line shows the OH1 model (see Sect.\,\ref{ssec_disc_uncert_dust}; 
   for clarity, only the absorption coefficient spectrum is shown). 
   Vertical dashed lines mark the wavelength range of the original OH models, i.e., values outside this range are extrapolated (see text).
   Arrows indicate the wavelength bands of the dust emission data used in this paper.}
   \label{fig_dustop}
\end{figure}

For the purpose of this paper, we adopted a dust opacity model from \citet{oh94}. Their models assume dust grains consisting of a 
mixture of silicates and amorphous carbon with different levels of coagulation and ice layer coverage around the agglomerates. 
Starting with the Mathis-Rumpl-Nordsieck (MRN) size distribution \citep{mathis1977}, the dust grains are processed (coagulated) within a certain time and at 
a certain density. We chose the model for moderately processed grains with a coagulation time of $10^5$\,yr at a density of $10^5$\,cm$^{-3}$\ 
and with thin ice mantles (hereafter called OH5a\footnote{We note that this opacity model is slightly different from the often-used 
OH5 model which assumes coagulation at a density of $10^6$\,cm$^{-3}$ (which is higher than what we actually observe in the cores). 
The 'OH5a' model is not tabulated in the \citet{oh94} paper, but is available on-line at \mbox{ftp://cdsarc.u-strasbg.fr/pub/cats/J/A+A/291/943.}}). 
These properties are closest to those typically observed in the starless cores in our sample \citep[see Tab.\,\ref{t_coreproperties} and][]{lippok2013}. 
The mass absorption coefficient of this dust model at $\lambda=1.2$\,mm is $\kappa_\mathrm{1.2mm}=0.79$\,cm$^2$\,g$^{-1}$\ of dust.
It is thus somewhere in the middle of the range of values covered by several other published dust models 
\citep[$\approx0.2-2$\,cm$^2$\,g$^{-1}$; e.g.,][]{oh94,weingartner2001,ormel2011}.
For converting the dust mass into hydrogen mass, we adopt a mean hydrogen-to-dust mass ratio in the solar neighborhood of 
\mbox{$M_{\rm H}/M_{\rm d}=110$} \citep[e.g.,][]{sodroski1997}. 
To obtain the total gas mass, \mbox{$M_{\rm g}$}, the hydrogen mass must be multiplied by the mean atomic weight per H nucleus of the ISM, 
for which we adopt a value of $\mu=1.4$\ \citep{cox2000}, i.e., \mbox{$M_{\rm g}/M_{\rm d}\approx150$}.

The radiative transfer calculations require dust opacity values over the wavelength range between 90\,nm and 10\,mm to sample the 
full spectrum of the ISRF (Sect.\,\ref{ssec_mod_srt} and Fig.\,\ref{fig_ISRF}).
The dust models of \citet{oh94}, however, cover only the wavelength range between 1\,$\mu$m and 1.3\,mm. 
Therefore, we extrapolate the dust opacities into the ultra-violet using the prescription of 
\citet[][ Eq. 1, Tab. 3, Col. 5]{cardelli1989}, and from 1.3 to 10\,mm using a simple power-law. 
Finally, since \citet{oh94} do not give scattering efficiencies, we augmented the OH5a model with 
scattering efficiencies following the approach of \citet{young2005} and using the albedos from the WD3.1 
model \citep{weingartner2001}.
The corresponding opacity spectrum is shown in Fig.\,\ref{fig_dustop}.
The albedos from the WD5.5B model \citep{weingartner2001} would probably be the more appropriate equivalent to the 
coagulated OH5a dust model, but given the impossibility of calculating scattering efficiencies for such complex dust models 
as \citet{oh94}, for this paper, we use the WD3.1 albedos and restrict ourselves to a discussion of the 
related uncertainties in Sect.\,\ref{ssec_disc_uncert_dust}.


\subsection{Ray-tracing inversion method}    \label{ssec_mod_rti}

The ray-tracing inversion technique derives a best-fit ($\chi^2$\ minimization) dust temperature and volume density distribution
to the uncertainty-weighted dust continuum flux density maps of molecular cloud cores. To work properly, flux density maps 
at three or more wavelengths, optimally distributed around the peak of the thermal SED, are required.
The method needs as input parameterized descriptions of the functional form of the density and temperature profiles, 
as well as a dust opacity model (Sect.\,\ref{ssec_mod_dust}). 
Since the profile parameter values are fit independently for each LoS, moderate deviations from spherical or elliptical 
symmetry can be easily reproduced. 
The technique was first described in \citet{nielbock2012} and initial results for individual starless cores from the EPoS sample 
were presented in \citet{nielbock2012}, \citet{lippok2013}, and \citet{schmalzl2014}. 
Here we only list the analytical profile formulas used in this method to illustrate the meaning of the profile parameters listed 
in Tab.\,\ref{t_coreproperties}.

\begin{figure}[htb]
   \centering
   \includegraphics[width=0.45\textwidth]{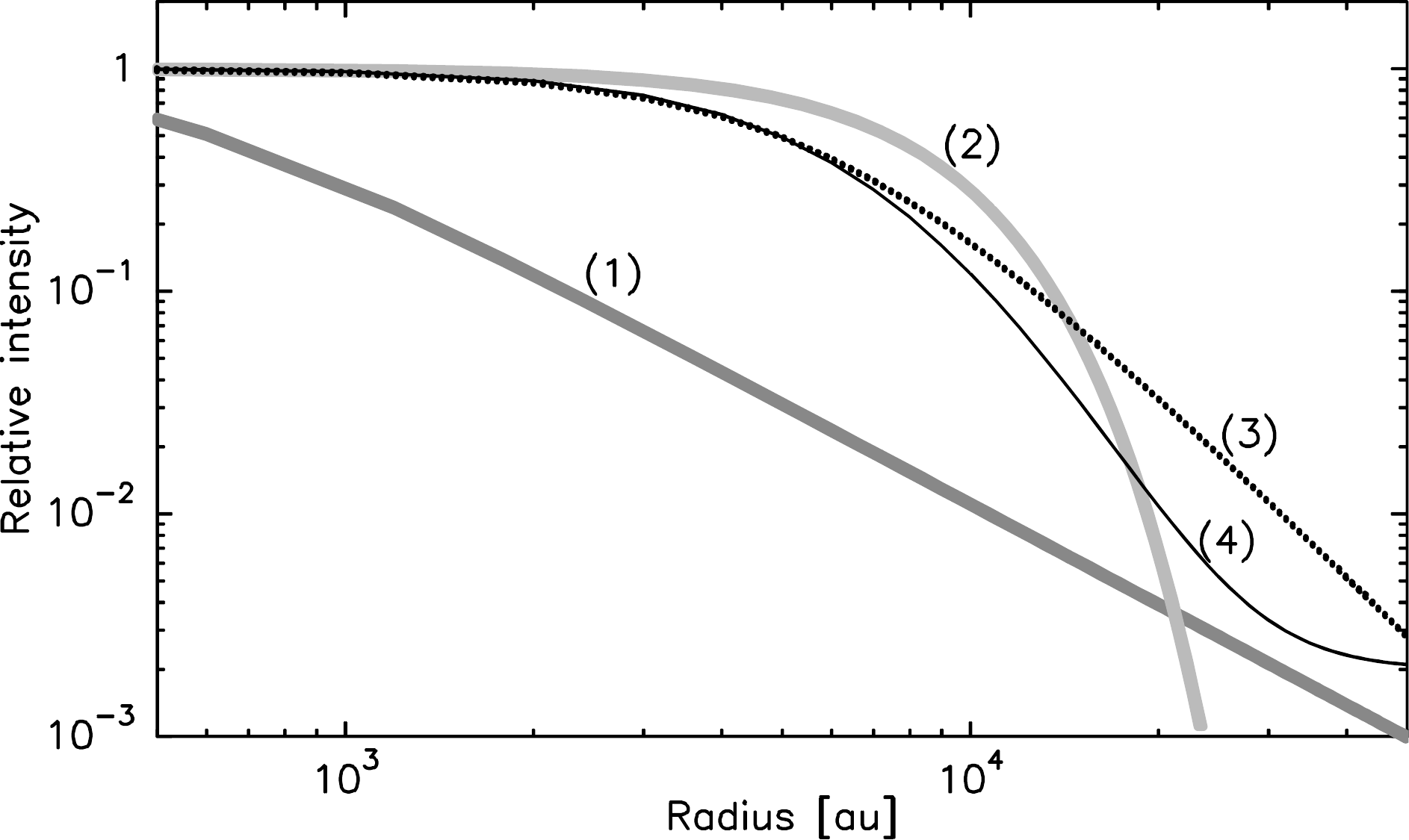}
   \caption{Modified normalized Plummer profiles (Eq.\,\ref{eq_plummer}) with parameters chosen to mimic
   (1) a simple power law, (2) a flat-density core with smooth outer edge, 
   (3) an isothermal BES, and (4) a profile like observed in B\,68.
   }
   \label{fig:plummer}
\end{figure}

For the density profile, we use a universal modified Plummer-like function, which 
can, for a given choice of parameters, mimic various kinds of profiles (Fig.\,\ref{fig:plummer}), 
including a simple (truncated) power-law, a nearly constant density sphere (with smooth edge), 
or a Bonnor-Ebert (BE) density profile 
\citep[][]{plummer1911,ebert1955,bonnor1956,whitworth2001,nielbock2012}.
This modified Plummer profile is described by
\begin{equation}
n_\mathrm{H}(r) = \frac{\Delta
  n}{\left(1+\left(\frac{r}{r_0}\right)^2\right)^{\eta/2}}+n_\mathrm{out} \;\;\mathrm{if} \; r\le r_\mathrm{out}
\label{eq_plummer}
\end{equation}
where $n_\mathrm H=2\,n(\mathrm H_2)+n(\mathrm H)$ is the total number density of hydrogen nuclei. 
This profile 
\textit{(i)}~accounts for an inner flat-density core inside $r_0$\ with a peak density 
$n_0=\Delta n+n_\mathrm{out}$\ (usually $n_\mathrm{out}\ll \Delta n$), 
\textit{(ii)}~approaches a power-law with exponent $\eta$\ at $r\gg r_0$,
\textit{(iii)}~turns over at \mbox{$r_2\approx\,r_0\,(\Delta n/n_\mathrm{out})^{1/\eta}$}\ into a flat-density halo approaching $n_\mathrm{out}$\,, and 
\textit{(iv)}~is cut off at $r_\mathrm{out}$. 
The tenuous envelopes, which we have shown in \citet{launhardt2013} to surround most globules and which become often evident 
as cloudshine\footnote{Stellar photons that are scattered at small dust grains in the optically thin halos \citep[e.g.,][]{foster2006}.}
halo, are actually neither azimuthally symmetric nor always fully spatially recovered by our observations of the dust emission. 
Therefore, $r_\mathrm{out}$\ is only estimated from the circularized 500\,$\mu$m emission profiles, 
which show a similar spatial extent as the cloudshine \citep[see Figs. A.1 through A.12 in][]{launhardt2013}.
The values of the free profile parameters $\Delta n$, $r_0$, $\eta$, and $n_\mathrm{out}$\ are iterated in the ray-tracing inversion 
until convergence is achieved between LoS (input) and PoS (plane of sky, output), as described in \citet{nielbock2012}.

\begin{figure}[htb]
   \centering
   \includegraphics[width=0.49\textwidth]{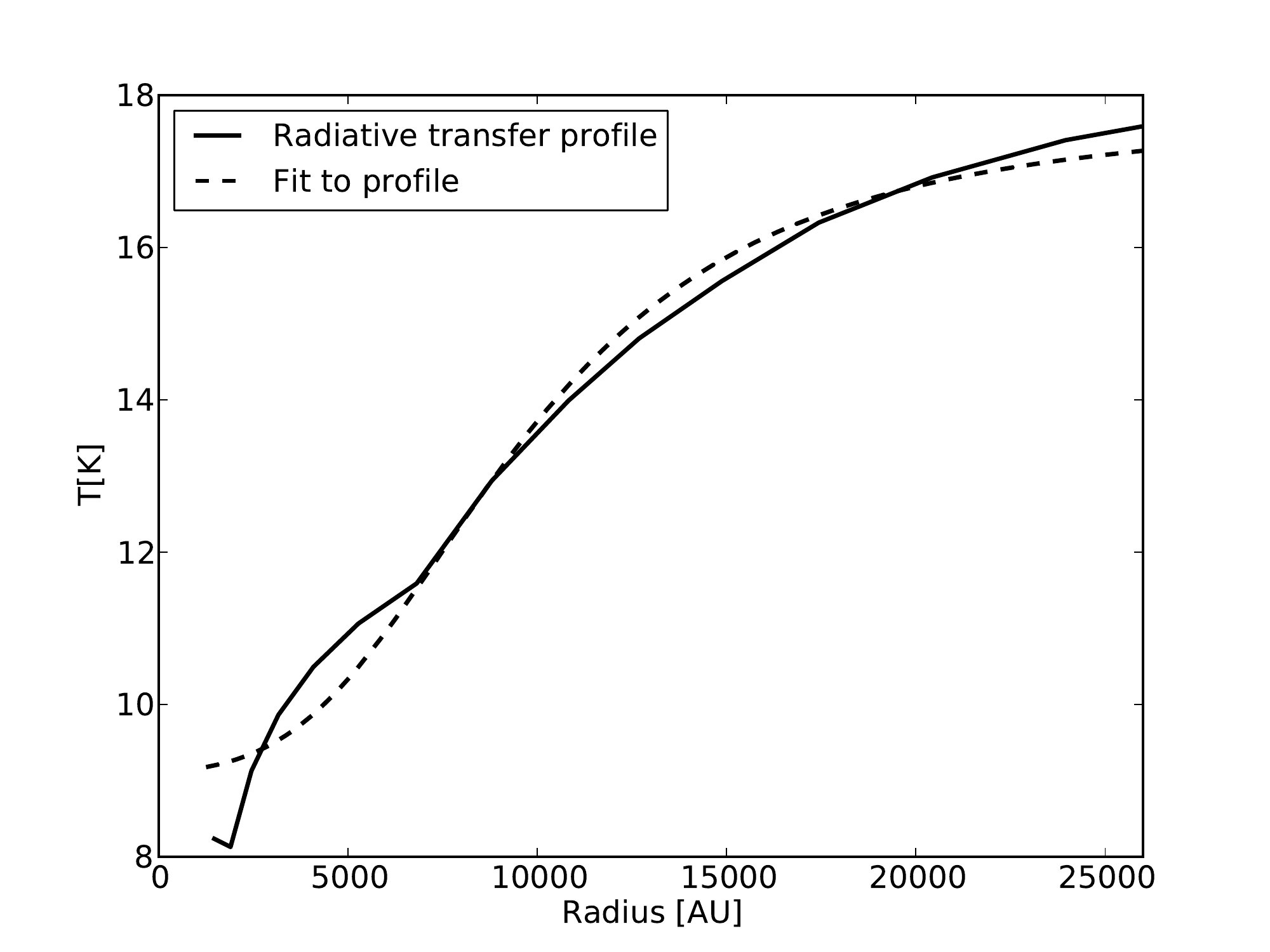}
   \caption{Comparison of the self-consistently calculated equilibrium temperature profile for B\,68 (solid line) 
          and an analytical profile fit following Eq.\,\ref{eq:T}. The little kinks in the radiative transfer temperature curve 
          at $r\approx 2000$\ and 7000\,au are computational artefacts due to the high optical depth.}
   \label{fig_comp_radTrans_profile}
\end{figure}

For the temperature structure of the starless cores, we use an empirical profile function that resembles the radiation 
transfer equation for an externally heated cloud. 
It couples the local temperature to the effective optical depth toward the outer 'rim' at which the ISRF impacts:
\begin{equation}
T(r)=T_\mathrm{out}-\Delta T\left(1-\mathrm{e}^{-\tau_\mathrm{ISRF}(r)}\right)
\label{eq:T}
\end{equation}
with $\Delta T = T_\mathrm{out}-T_\mathrm{min}$~and the frequency-averaged effective optical depth
\begin{equation}
\tau_\mathrm{ISRF}(r)=\tau_0 \frac{\int_{r}^{r_\mathrm{out}} n_\mathrm{H}(x)\,\mathrm{d}x}{\int_{0}^{r_\mathrm{out}}n_\mathrm H(x)\,\mathrm d x}.
\end{equation}
$\tau_0$\ is an empirical (i.e.~free) scaling parameter that accounts for the a priori unknown mean dust opacity and the
SED of the UV radiation of the ISRF. The temperature at the core center, 
$T_0\equiv T(r=0)$, converges to $T_\mathrm{min}$\ if $\tau_\mathrm{ISRF}\gg 1$.
As for the density profile, the values of the free profile parameters, $T_\mathrm{out}$, $\Delta T$, and $\tau_0$, 
are iterated until convergence is reached between LoS input and PoS output.
Figure\,\ref{fig_comp_radTrans_profile} compares the equilibrium temperature profile for B\,68, self-consistently modeled with radiative transfer,
with an analytical profile fit using Eq.\,\ref{eq:T}. The good agreement between the two profiles suggests that Eq.\,\ref{eq:T} provides a
reasonable parameterization of the temperature profile of an externally heated starless core.

We also tested the uniqueness of the ray-tracing solution in $\chi^2$\ space. While the solution 
was found to be genuinely unique for most pixels and most density/temperature profiles, the confidence regions can 
become quite elongated (banana-shaped) and even break up into two or three minima for certain configurations and pixels, 
in particular in regions with steep density and temperature gradients \citep[see also][for a discussion of this problem in a slightly different parameter space]{juvela2012}.
However, the occasional jumping of the solution to one of these secondary (unphysical) 
minima can be easily circumvented by implementing a very low-weighted near-infrared (NIR) 
extinction map (or any other smooth column density proxy), which effectively suppresses the secondary minima in the $\chi^2$\ plane
without affecting the value of the primary solution. 
Thus, the solution is always unique in practice.
Moreover, the relative smoothness of the resulting mid-plane dust temperature and integrated column density maps (Fig.\,\ref{fig-tmaps}),
which are composed of independently fit map pixels with different wavelength coverage (due to different map sizes at different wavelengths)
also shows that the solution of the ray-tracing inversion is stable against both noise in the individual flux maps as well as 
wavelength coverage (but see wavelength coverage requirement mentioned above).


\subsection{Self-consistent radiative transfer modeling}   \label{ssec_mod_srt}

We use the radiative transfer code HYPERION \citep{robitaille2011} to self-consistently calculate the equilibrium 
temperature distributions of the starless cores. As inputs we use the 
the azimuthally averaged radial density profiles derived with the ray-tracing 
inversion technique and the same OH5a dust opacity model. The only free parameter in the modeling is  
$s_\mathrm{ISRF}$, the relative total strengths of the ISRF.
In addition, we allow $N_\mathrm{H}(r_\mathrm{sym})$, the column density of the surrounding envelope 
that shields the core from the ISRF, to vary within its observational constraints (see Tab.\,\ref{t_coreproperties}).
The best-fit values of  $s_\mathrm{ISRF}$\ and 
$N_\mathrm{H}(r_\mathrm{sym})$\ are then determined by comparing the calculated temperature profiles with the ray-tracing 
results as described in Sect.\,\ref{ssec_mod_comp}.

In our physical model for the radiative transfer, the dust is heated by the ISRF alone and cools by emitting thermal radiation.
Additional heating via, for example, collisions with molecules, which are themselves heated by cosmic rays, 
is neglected since previous studies showed that this effect is very weak even at the highest densities found 
in the starless cores considered here \citep[$<1.5$\,K,][]{goldsmith2001,evans2001}.
Heating by a potential compression of the cores is also neglected since this becomes relevant only at lower 
temperatures and higher densities than present in our cores \citep{keto2010}.

Since we only compare radial profiles and carry out 1-D radiative transfer calculations, the model cores are spherically 
symmetric and the ISRF is considered isotropic. 
However, the (approximate) spherical symmetry of the actual cores breaks down at large radii in 
most cases \citep[see Figs.\,A1 through A12 in][]{launhardt2013},
such that the 1-D approximation may no longer hold even with azimuthal averaging. 
We therefore determine (visually) and list in Table\,\ref{t_coreproperties}
for each core a radius $r_\mathrm{sym}$\ inside which the core can safely be considered spherically symmetric.
Temperature profiles from the two methods are only compared inside $r_\mathrm{sym}$. 
To still account for the attenuation of the ISRF by the material outside of $r_\mathrm{sym}$, we 
estimate the mean attenuation of the ISRF at $r_\mathrm{sym}$\ from the observed LoS column density, 
$N_\mathrm{H}(r_\mathrm{sym})$, assuming spherical symmetry.
The uncertainty range of $N_\mathrm{H}(r_\mathrm{sym})$, 
which is mostly related to the actual deviations from spherical symmetry of the envelope and 
which we derive from the azimuthal scatter of  $N_\mathrm{H}$\ values at $r_\mathrm{sym}$, 
is accounted for by varying $r_\mathrm{out}$\ in Eq.\,\ref{eq_plummer}. The observed values and uncertainty ranges of 
$N_\mathrm{H}(r_\mathrm{sym})$\ are also listed in Table\,\ref{t_coreproperties}.

\begin{figure}[htb]
   \centering
   \includegraphics[width=0.49\textwidth]{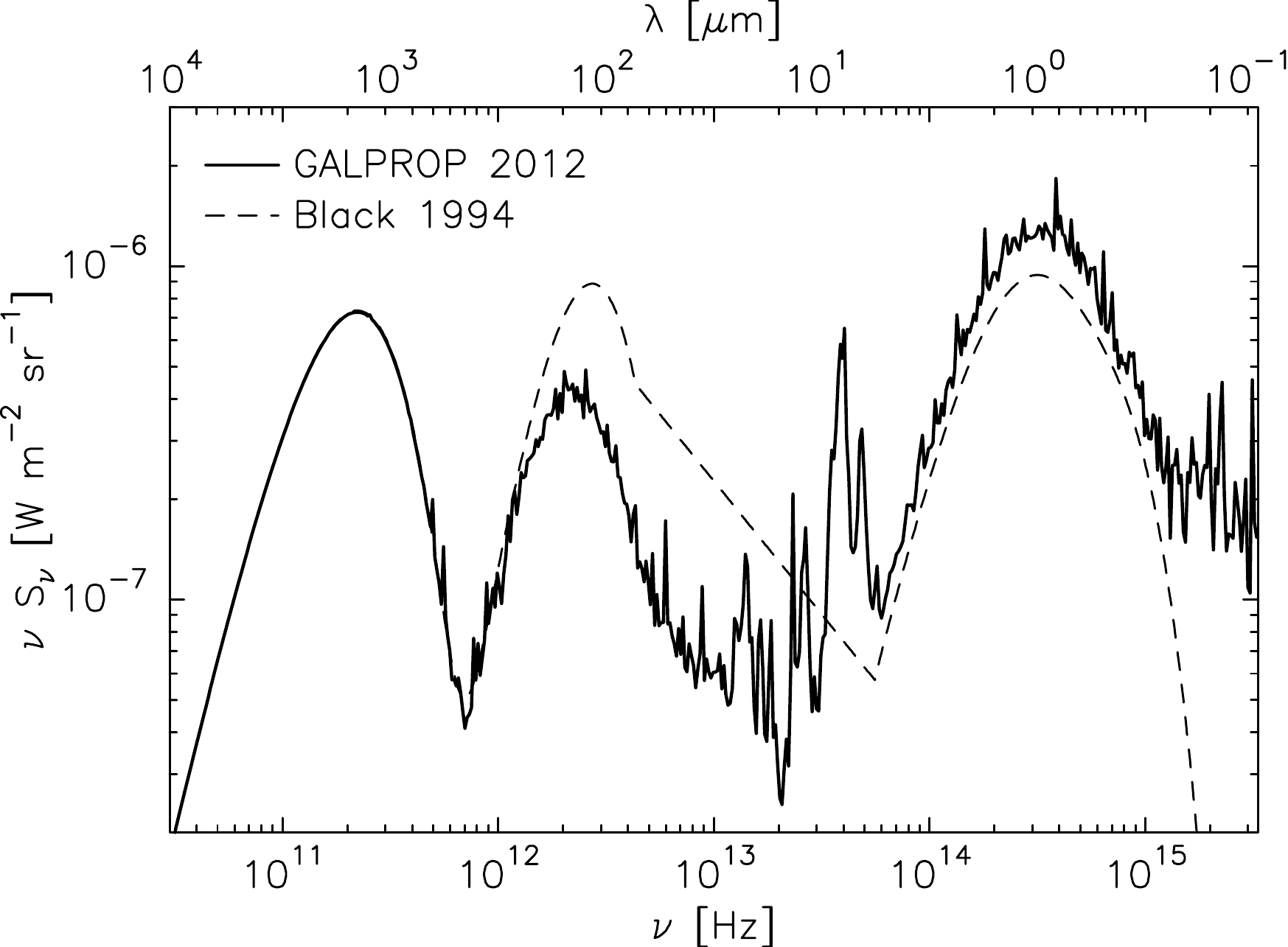}
   \caption{ISRF model developed by the GALPROP consortium for the solar neighbourhood \citep{ackermann2012} 
   with the CMB already included (solid line). For comparison we also show the model by 
   \citet{black1994} as parameterized in \citet{zucconi2001}.}
   \label{fig_ISRF}
\end{figure}

For the ISRF, we use a model that was developed by the GALPROP consortium \citep{ackermann2012}.
From their four-component model for the Milky Way (stellar, scattered, transient, thermal), 
we obtained a table with the SED at the grid position corresponding to the 
Sun's location in the Milky Way ($R_\mathrm{GC}=8.5$\,kpc).
Added to that model is the cosmic microwave background (CMB) as a 2.72\,K black-body.
Figure\,\ref{fig_ISRF} shows the SED of our adopted ISRF model, along with that of the "classical" model by 
\citet{black1994}.
Compared to this older model, the GALPROP model has a $\approx$1.5 times higher flux at wavelengths shortward of 8\,$\mu$m 
(mostly stellar contribution), which is the most important spectral range for dust heating.
Both the total and relative contributions of the different components in the GALPROP ISRF model have 
uncertainties that are difficult to quantify.
Furthermore, individual components of the ISRF may vary differently locally on spatial scales smaller than sampled by the model, 
for example, due to the proximity to luminous stars, to star-forming regions, or to molecular clouds that 
can also shield a region from the general ISRF (not only for locations inside the cloud, but also if a dark cloud 
is located between a globule and the galactic plane). 
Since we have no means of constraining such local variations of the individual components 
better than done in the GALPROP model, 
we took the simplifying approach to account for the uncertainty of the local ISRF 
by introducing the free scaling parameter $s_\mathrm{ISRF}$, which multiplies the total strength 
of the GALPROP ISRF (i.e., excluding the CMB).
In Sect.\,\ref{ssec_disc_uncert_isrf}, we also test a different spectral shape of the ISRF and discuss its effect on the results.


\subsection{Comparing ray-tracing and self-consistent radiative transfer results}   \label{ssec_mod_comp}

With the setup described above, we calculated radiative transfer models for grids of the two free parameters 
$s_\mathrm{ISRF}$\ and $N_\mathrm{H}(r_\mathrm{sym})$. The latter is only free within the observational constraints.
The value of $s_\mathrm{ISRF}$\ was varied in 19 steps between 0.1 and 5.0 for all sources. 
This total range is somewhat larger than the range of ISRF scaling factors listed by \citet{ackermann2012}
for different cosmic ray models and different locations in the Galaxy. Since all our sources are located in the 
Solar neighbourhood, are relatively isolated, and are not located close to star-forming regions or luminous 
individual stars, it should thus cover the uncertainty in our knowledge of the local strength of the ISRF.

The total number of models per source depended on the uncertainty range 
of $N_\mathrm{H}(r_\mathrm{sym})$\ (Sect.\,\ref{ssec_res_rti} and Tab.\,\ref{t_coreproperties}) 
and ranged from 57 to 209.
The best-fit values of $s_\mathrm{ISRF}$\ and $N_\mathrm{H}(r_\mathrm{sym})$\
(which we label $s^{\prime}_\mathrm{ISRF}$\ and $N^{\prime}_H(r_\mathrm{sym})$) are then determined 
by means of $\chi^2$\ minimization of the differences between the temperature profiles at $r\le r_\mathrm{sym}$\
predicted by the radiative transfer models and those derived with the ray-tracing inversion. 
In addition to deriving the best-fit parameter values of  $s_\mathrm{ISRF}$\ and $N_\mathrm{H}(r_\mathrm{sym})$,
we evaluate the goodness of the fits by verifying if the best-fit equilibrium temperature distributions 
agree with the temperatures of the ray-tracing inversion technique within its uncertainty of $\sigma_T=^{+2}_{-1}$\,K 
(see Sect.\,\ref{ssec_disc_uncert}) everywhere within $r_\mathrm{sym}$.


\section{Modeling results}    \label{sec_res}


\subsection{Ray-tracing results}    \label{ssec_res_rti}

\begin{table*}[htb]
\caption{\label{t_coreproperties}
Properties of the studied cores}
\centering
\begin{tabular}{lcccccccccccc}
\hline\hline
Object                                                          & 
$n_0$\tablefootmark{a}                           & 
$n_{\mathrm out} $\tablefootmark{a}     & 
$N_0$\tablefootmark{a}                           & 
$r_\mathrm{0}$\tablefootmark{a}           & 
$r_\mathrm{out}$\tablefootmark{a}       & 
$r_\mathrm{sym}$                                    &
$N_H(r_\mathrm{sym})$                         &
$\eta$\tablefootmark{a}                           &
$T_0$\tablefootmark{a}                           & 
$T_\mathrm{out}$\tablefootmark{a}      & 
$M_\mathrm{core}$\tablefootmark{b}   & 
$L_\mathrm{bol}$\tablefootmark{c}      \\
 & [cm$^{-3}$] &  [cm$^{-3}$] & [cm$^{-2}$] &  [au] & [au] & [au] & [cm$^{-2}$] & & [K] & [K] & [$M_{\sun}$] & [$L_{\sun}$] \\
\hline
CB\,4\,-\,SMM\tablefootmark{1,3}      & 2.5E4 & 1E1 & 1.0E22 & 1.7E4 & 9.0E4 & 4.1E4 & (0.4$\pm$0.2)E21 & 5.2 & 11.9 & 19.0 &  2.9  & 0.77 \\
CB\,17\,-\,SMM\tablefootmark{2,4}    & 1.1E5 & 6E2 & 3.0E22 & 1.2E4 & 5.5E4 & 2.8E4 & (3.5$\pm$1.0)E21 & 4.8 & 10.4 & 13.5 &  3.9  & 0.37 \\
CB\,26\,-\,SMM2\tablefootmark{1,5}  & 8.0E4 & 5E2 & 1.7E22 & 7.5e3 & 3.9E4 & 2.5E4 & (2.5$\pm$1.0)E21 & 3.0 & 11.8 & 14.2 &  3.1  & 0.30 \\
CB\,27\,-\,SMM\tablefootmark{2,3}    & 1.1E5 & 1E3 & 2.6E22 & 1.1E4 & 4.2E4 & 2.9E4 & (5.0$\pm$1.5)E21 & 4.0 & ~9.8 & 14.4 &  6.0  & 0.44 \\
B\,68\,-\,SMM\tablefootmark{1,3}       & 1.6E5 & 3E2 & 2.8E22 & 1.0E4 & 4.7E4 & 2.7E4 & (1.0$\pm$0.5)E21 & 5.0 & ~7.7 & 17.3 &  2.6  & 0.23 \\
CB\,244\,-\,SMM2\tablefootmark{1,4} & 3.2E5 & 5E2 & 9.2E22 & 6.3E3 & 7.2E4 & 3.5E4 & (5.0$\pm$2.0)E21 & 2.6 & ~7.5 & 14.0 & 14  & 0.53 \\
\hline
\end{tabular}
\tablefoot{
\tablefoottext{a}{Values determined from circular profile fits to the results of the ray-tracing inversion technique using the OH5a dust model; 
 see Sect.\,\ref{ssec_mod_rti}.}
\tablefoottext{b}{Derived by integrating the column density maps out to $R=5\times10^4$\,au (see Fig.\,\ref{fig_massradius})
                             and corrected for He and metals with $\mu=1.4$\ (see Sect.\,\ref{ssec_mod_dust}).}
\tablefoottext{b}{Taken from \citet[][Table\,6]{launhardt2013}.}
\newline Source-related remarks \citep[see also][]{launhardt2013}:
\tablefoottext{1}{Relatively round or only slightly elliptical core.}
\tablefoottext{2}{Very elliptical core or pronounced tail.}
\tablefoottext{3}{Single core and no nearby protostar.}
\tablefoottext{4}{Multiple cores and nearby protostar with partial blending.}
\tablefoottext{5}{Not a coherent core but super-projection of 2-3 filamentary cores.}
}
\end{table*}

\begin{figure*}[htbp]
 \centering
  \includegraphics[width=0.95\textwidth]{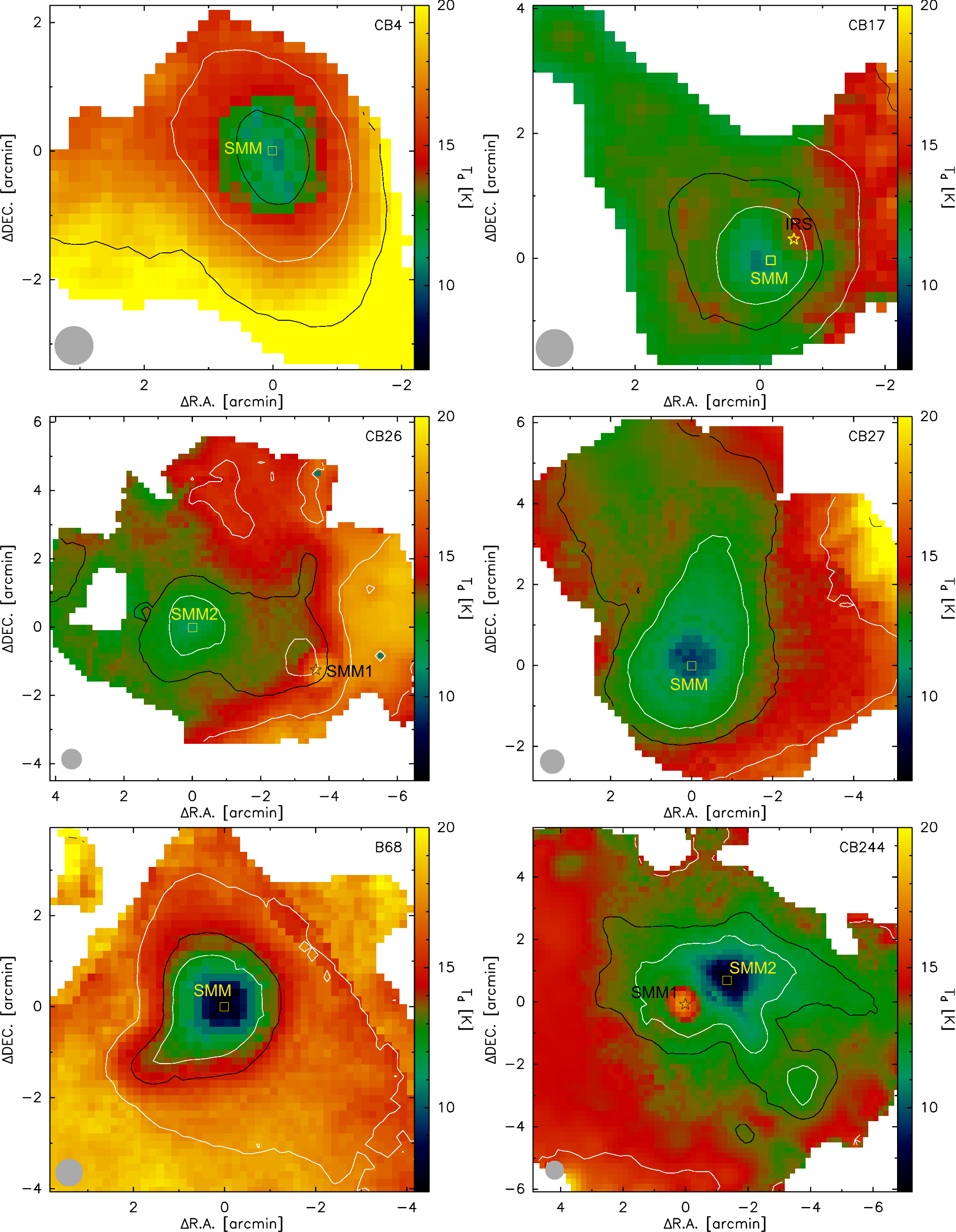}
 \caption{\label{fig-tmaps}
  Mid-plane dust temperature maps (color, all maps on the same scale) of all six globules studied in this paper,
  derived with the ray-tracing inversion technique and  
  overlaid with contours of the hydrogen column density (white: $10^{21}$\ and $10^{22}$\,cm$^{-2}$, 
  black: 0.5 and $5\times10^{21}$\,cm$^{-2}$).
  Yellow squares indicate the center positions of the starless cores as listed in Tab.\,\ref{t1} and in 
  \citet{launhardt2013}. Asterisks indicate the positions of embedded protostars \citep[see][]{launhardt2013}.}
\end{figure*}

\begin{figure*}[htb]
   \centering
   \includegraphics[width=0.95\textwidth]{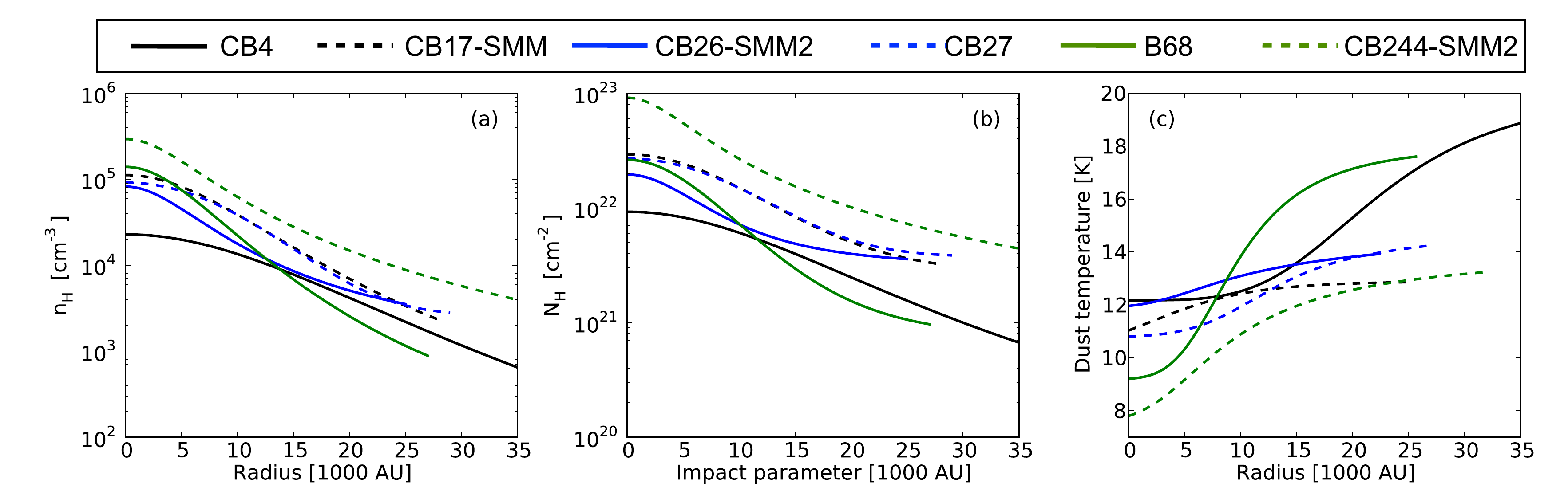}
   \caption{Radial profiles of all six starless cores, derived with the ray-tracing inversion technique and plotted up to $r_\mathrm{sym}$. 
    (a)~mid-plane hydrogen volume number density vs. radius,
    (b)~hydrogen column density vs. impact radius, 
    (c)~mid-plane dust temperature vs. radius.}
   \label{fig_dustprofiles}
\end{figure*}

We derived the dust temperature and density structure of all six starless cores
using the ray-tracing inversion technique and the OH5a dust opacity model to fit the dust continuum flux density 
maps in up to eight bands between 100\,$\mu$m and 1.2\,mm. 
The initially resulting maps of mid-plane dust temperature and density were already shown in \citet{lippok2013}.
Like in \citet{launhardt2013} and in \citet{lippok2013}, we derive a strong negative correlation between 
central dust temperature and peak column density, which suggests that the thermal structure of the cores 
is dominated by external heating from the ISRF and shielding by dusty envelopes.

For the present paper, we recalculated the temperature and density maps 
using an improved convergence scheme for the ray-tracing algorithm. 
This resulted in slightly different profile parameters from those of \citet[][Table\,3]{lippok2013},
albeit qualitatively and quantitatively very similar dust temperature and density maps.
The new values of the profile parameters are, however, within the formal uncertainties of the \citet{lippok2013} results 
and the differences are thus not significant.
The respective new mid-plane dust temperature maps and contours of the (integrated) column density are shown in Fig.\,\ref{fig-tmaps}.
The azimuthally averaged radial profiles of the mid-plane volume density and dust temperature as well as of the integrated column density 
are shown in Fig.\,\ref{fig_dustprofiles}. 
The corresponding radial profile fit parameters according to Eqs.\,\ref{eq_plummer} and \ref{eq:T} are listed in Table\,\ref{t_coreproperties}.
Note that $\eta$\ (eq.\,\ref{eq_plummer}) is only listed for completeness here; its value has a large uncertainty since it is sensitive 
to how the relative weights of individual data points (pixels) in the profile fitting are scaled with radius and signal-to-noise ratio
(see also uncertainty discussion in Sect.\,\ref{ssec_disc_uncert_dust}).
Only these radial profiles are used as input for and comparison with the self-consistent 1-D radiative transfer modeling.

\begin{figure}[htb]
   \centering
   \includegraphics[width=0.45\textwidth]{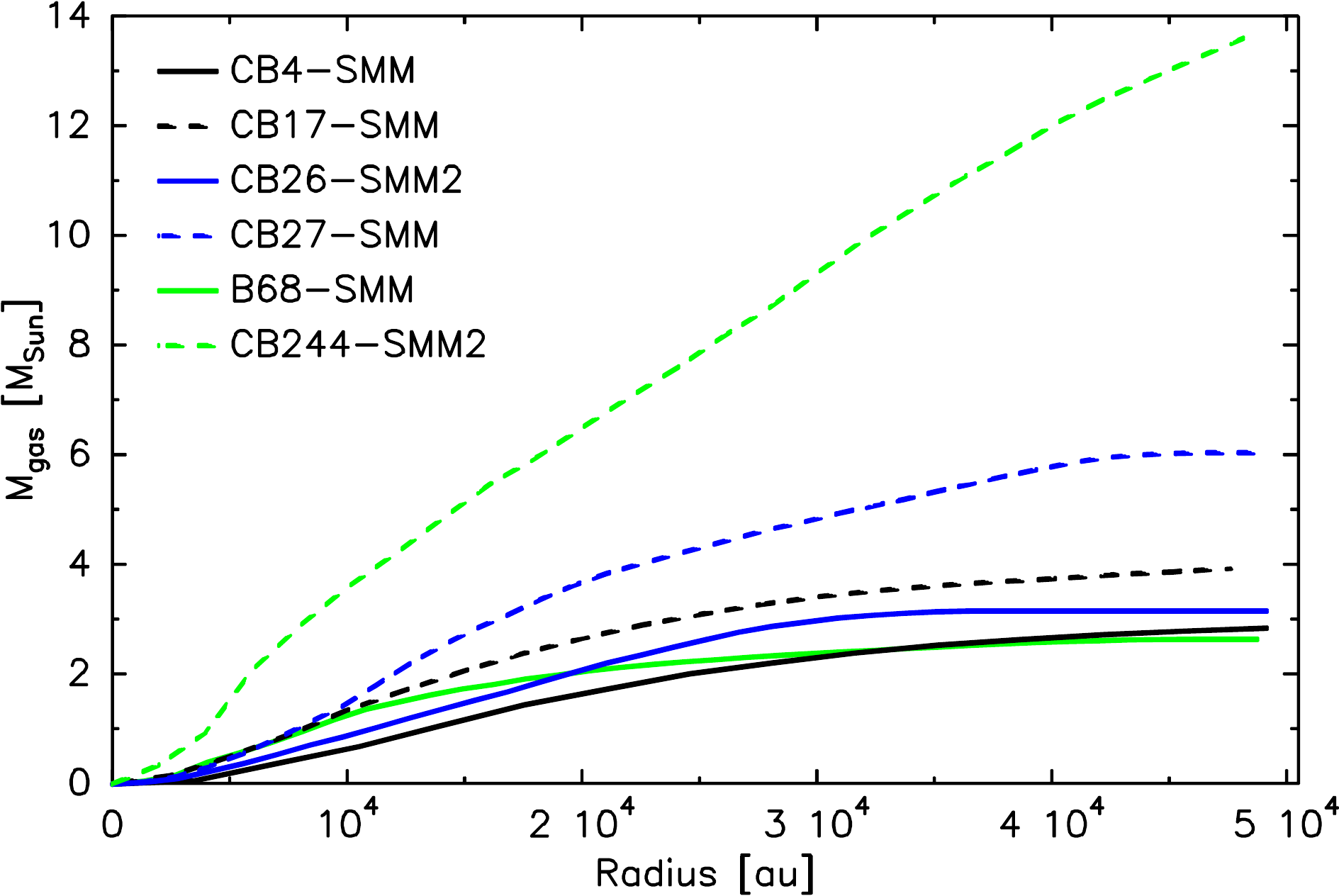}
   \caption{Cumulative radial mass distribution of the six starless cores derived with the ray-tracing inversion technique 
   and after masking out the nearby protostars in CB\,17, CB26, and CB\,244.}
   \label{fig_massradius}
\end{figure}

Figure\,\ref{fig_massradius} shows the cumulative radial gas mass distributions of the six cores, 
derived from the column density maps after masking the overlapping protostars and their respective envelopes in 
CB\,17, CB\,26, and CB\,244. 
The size of masking regions was chosen to be a compromise between avoiding the local surroundings 
of the warm protostars and using as much as possible the extended emission from the cold cores since the 
protostellar envelopes neither have sharp boundaries nor can they clearly be separated in the emission maps.

The physical outer radii of the globules are all within the narrow range $(6.5\pm2.5)\times10^4$\,au (Table\,\ref{t_coreproperties}),
and, with the exception of CB\,244, the cumulative mass distributions are basically flat beyond $r\approx4.5\times10^4$\,au 
(Fig.\,\ref{fig_massradius}). This means that the more extended envelopes, which are partially visible as cloudshine 
\citep[see][Figs.\,A.1 through A.12]{launhardt2013}, do not contribute much to the total mass. 
To ensure good comparability of the sources, we therefore chose to derive the total source masses from the cumulative 
mass distributions (Fig.\,\ref{fig_massradius}) within a fixed radius of $5\times10^4$\,au around the column density peaks of the starless cores. 
The resulting total gas masses are listed in Table\,\ref{t_coreproperties} and are in the range 2.6\,--\,14\,M$_{\odot}$.

\begin{figure}[htb]
   \centering
   \includegraphics[width=0.45\textwidth]{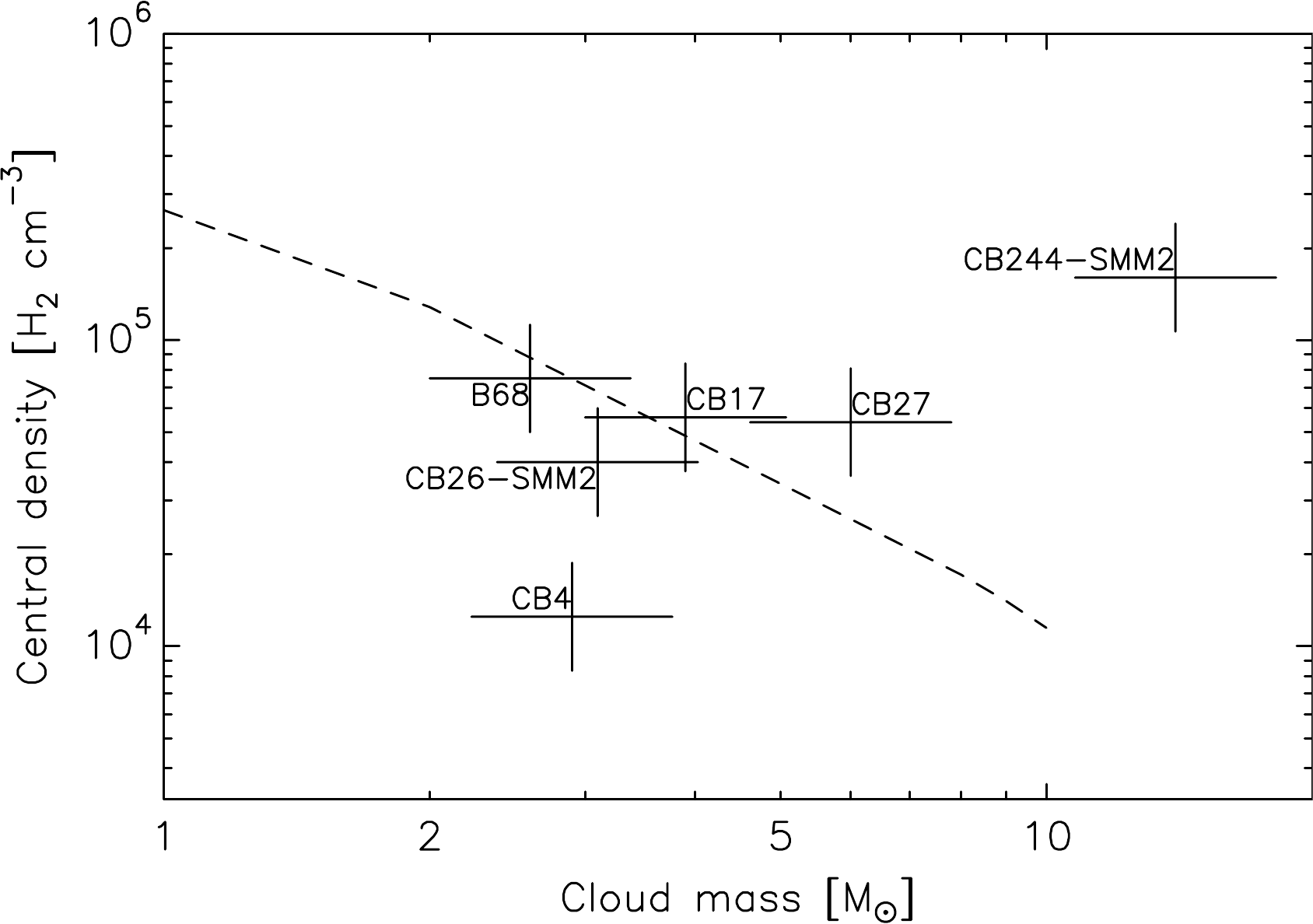}
   \caption{Central density vs. total gas mass for the six starless cores (data from Table\,\ref{t_coreproperties}).
   The dashed line indicates the maximum stable density of a pressure-supported, self-gravitating modified (nonisothermal) 
   BES as calculated by \citet[][their model with photoelectric heating at the core boundary taken into account]{keto2008}. 
   Error bars represent the mean relative uncertainties of 30\% on the cloud mass and 50\% on the central density (see Sect.\,\ref{ssec_disc_uncert}).
   }
   \label{fig_massdens}
\end{figure}

To better characterize the cores and to help interpreting the results, we revisited the stability assessment 
of \citet{launhardt2013} and \citet{lippok2013}. 
Figure\,\ref{fig_massdens} shows the central H$_2$\ volume density vs. the total gas mass for the six starless 
cores (see Table\,\ref{t_coreproperties}) along with the stability criterion by \citet{keto2008} for pressure-confined, 
self-gravitating modified (non-isothermal) Bonnor-Ebert spheres \citep[BES,][]{ebert1955,bonnor1956}. 
CB\,17, CB\,26, CB\,27, and B\,68 are located close to the boundary between stable and unstable, making their 
evolutionary path unpredictable.
CB\,4 is clearly a subcritical core and must be purely pressure-confined (or transient).
CB\,244 (SMM2) is a clearly supercritical core and thus a good place to look for infall signatures. 
These conclusions agree well with the analysis of \citet{launhardt2013}.


\subsection{Radiative transfer results}     \label{sec_res_srt}

We have already shown in Sect.\,\ref{ssec_mod_rti} and Fig.\,\ref{fig_comp_radTrans_profile} that 
using Eq.\,\ref{eq:T} for the temperature distribution in the ray-tracing inversion leads, in principle, to qualitatively and 
quantitatively very similar temperature profiles derived with this technique and predicted by the radiative transfer models.
Before comparing both model results in more detail for the individual cores in Sec.\,\ref{ssec_res_comp} and discussing 
the uncertainties and implications in Sect.\,\ref{sec_disc}, we demonstrate here the systematic effects of varying the two free parameters used 
in the radiative transfer modeling, $s_\mathrm{ISRF}$\ and $N_\mathrm{H}(r_\mathrm{sym})$.

\begin{figure}[htb]
   \centering
   \includegraphics[width=0.49\textwidth]{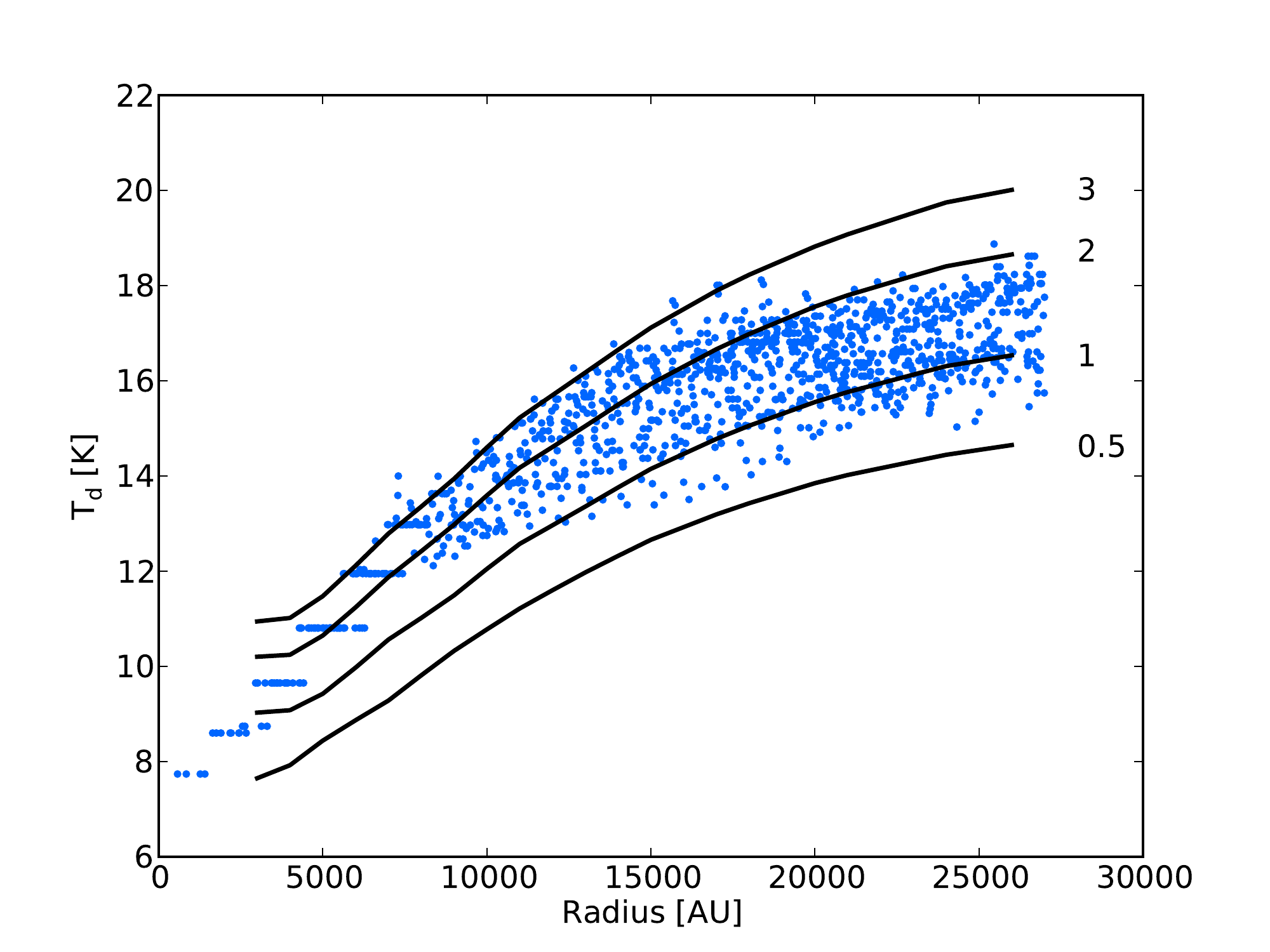}
   \caption{Effect on the radial equilibrium temperature distribution predicted for B\,68 
   from scaling the total flux of the GALPROP ISRF by factors $s_\mathrm{ISRF}=0.5,1,2,3$\
   (with OH5a dust and $N_\mathrm{H}(r_\mathrm{sym})=const=1.5\times10^{21}$\,H\,cm$^{-3}$; black lines). 
   Blue dots show the mid-plane dust temperature distribution inferred with the ray-tracing inversion.}
   \label{fig_b68_field}
\end{figure}

\begin{figure}[htb]
   \centering
   \includegraphics[width=0.49\textwidth]{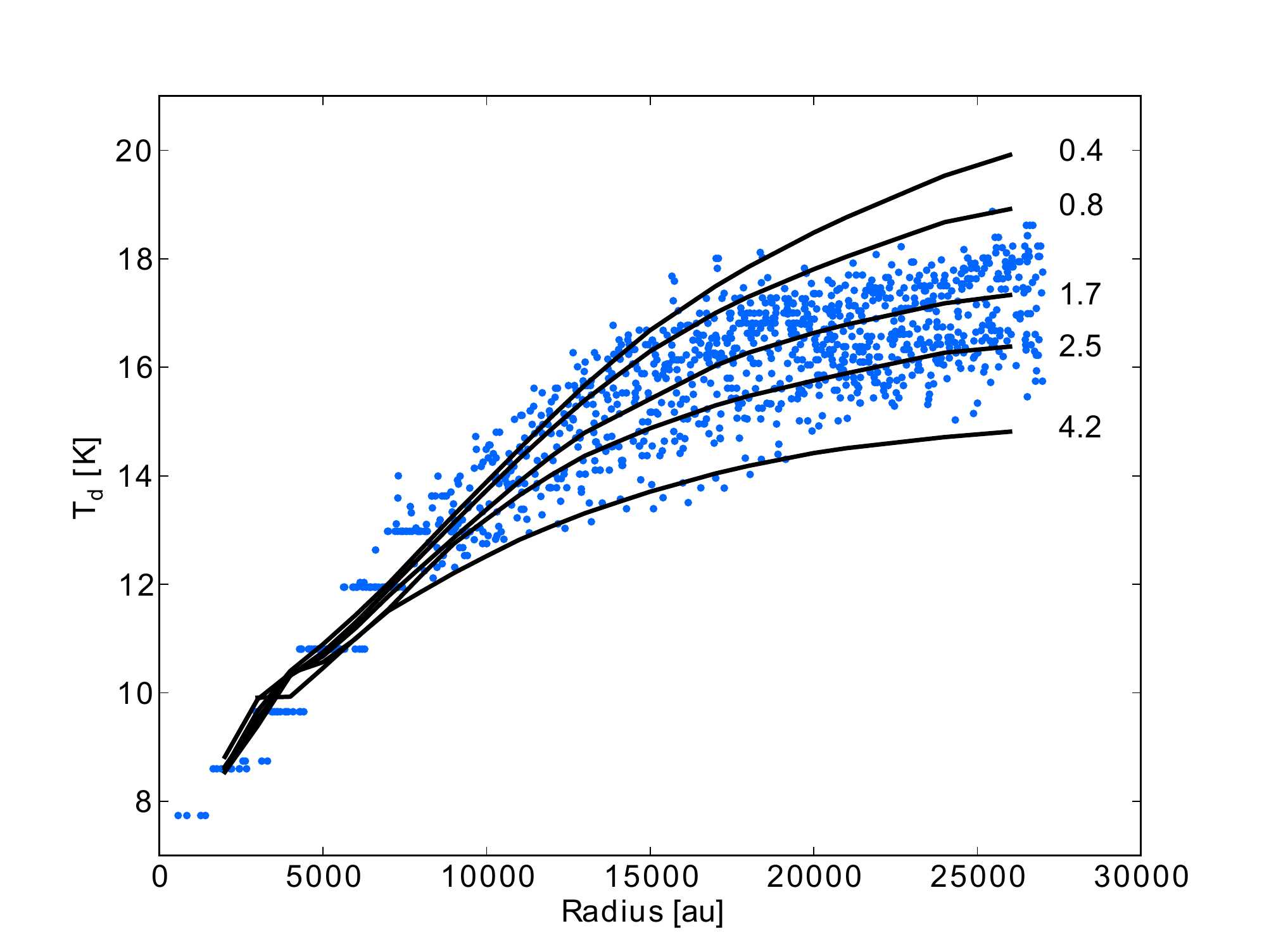}
   \caption{Effect on the radial equilibrium temperature distribution predicted for B\,68 
    from varying the envelope extinction of B\,68 via $N_\mathrm{H}(r_\mathrm{sym})$\, 
   for values of $0.4, 0.8, 1.7, 2.5, 4.2\times10^{21}$\,H\,cm$^{-3}$\
   (with OH5a dust and $s_\mathrm{ISRF}=const=2.2$; black lines). 
   Blue dots show the mid-plane dust temperature distribution inferred with the ray-tracing inversion.}
   \label{fig_b68_Ak}
\end{figure}

Figure\,\ref{fig_b68_field} compares the equilibrium temperature distributions in B\,68 for different ISRF scaling factors 
$s_\mathrm{ISRF}$\ (with OH5a dust and $N_\mathrm{H}(r_\mathrm{sym})=const=1.5\times10^{21}$\,H\,cm$^{-3}$). 
We find that the total strength of the ISRF scales the temperature approximately equally everywhere throughout the entire core. 
Figure\,\ref{fig_b68_Ak} compares temperature profiles of B\,68 for different envelope extinctions
(now leaving $s_\mathrm{ISRF}=const=2.2$).
In contrast to varying the total power of the ISRF, 
we find that changing the envelope extinction mainly affects the temperature at large radii only,
while the effect on the core temperature is marginal.
The reason for this behavior is that the outer layers mainly absorb the {\it UV} part of the ISRF,
while the radiation at longer wavelengths, that heats the interiors, is nearly unaffected.
Therefore, the temperature profiles for different envelope extinctions, but constant $s_\mathrm{ISRF}$, 
converge towards the core centers.
Hence, for a given dust model and spherical geometry, envelope extinction (or variations in the UV-to-IR flux ratio of the ISRF) 
controls the core-envelope temperature contrast, whereas scaling the entire ISRF mainly affects the mean temperature of the entire globule.


\subsection{Comparison of ray-tracing and radiative transfer results}    \label{ssec_res_comp}

\begin{figure*}[htb]
   \centering
   \includegraphics[width=0.95\textwidth]{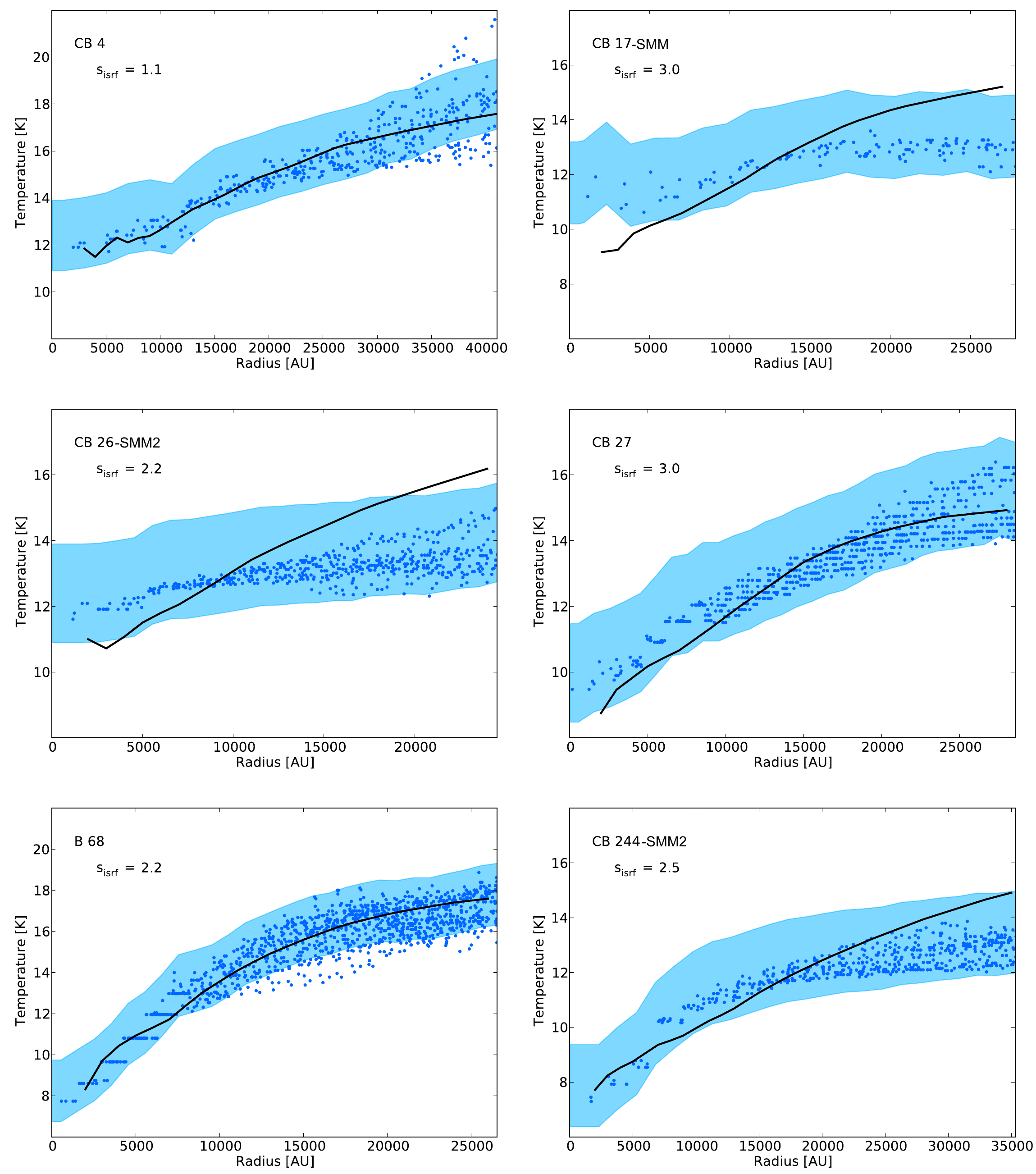}
   \caption{Comparison of the temperature distributions derived with the ray-tracing inversion technique (blue dots and shaded regions) 
   to the best-matching thermal equilibrium temperature distributions (black lines). 
   Blue dots represent pixel values of the mid-plane temperature map derived with the 
  ray-tracing inversion method. Shaded regions indicate the uncertainty of the mean temperature profile 
  which we estimate as $^{+2}_{-1}$\,K (Sect.\,\ref{ssec_disc_uncert}). 
  All models assume OH5a dust. 
  Best-matching scaling factors $s_\mathrm{IRSF}$\ for the total flux of the ISRF are also indicated in each panel.}
   \label{fig_bestmodels}
\end{figure*}

All six cores show a positive radial temperature gradient with cool interiors and warmer envelopes, 
as expected and predicted by radiative transfer models for clouds that are externally heated by the ISRF and shielded by dust.
This general behavior is readily reproduced for all sources by the ray-tracing inversion of the dust emission maps
(Fig.\,\ref{fig_dustprofiles}c).
On a more quantitative side, both the central core temperatures of 7.5\,--\,12\,K and the outer envelope temperatures 
of 13.5\,--\,19\,K determined with the ray-tracing method (Table\,\ref{t_coreproperties}) are well within the ranges that 
are predicted by self-consistent radiative transfer models (Fig.\,\ref{fig_bestmodels}).

\begin{table*}[htb]
\caption{\label{t_overview_results_1}
Best-fit values of $s_\mathrm{ISRF}$\ and $N_H(r_\mathrm{sym})$.}
\centering
\begin{tabular}{lccc}
\hline\hline
Object                                                                                  & 
$s^{\prime}_\mathrm{ISRF}$\tablefootmark{a}            & 
$N^{\prime}_H(r_\mathrm{sym})$\tablefootmark{a}   &
$\chi^2_\mathrm{red,min}$\tablefootmark{b}               \\
 & & [cm$^{-2}$] & \\
\hline
CB\,4\,-\,SMM       & $1.1\pm0.3$ & ($4.2\pm2.0$)E20 & 0.08 \\
CB\,17\,-\,SMM    & $3.0\pm0.5$ & ($4.5\pm1.0$)E21 & 1.10 \\ 
CB\,26\,-\,SMM2   & $2.2\pm0.3$ & ($3.0\pm0.5$)E21 & 0.80 \\
CB\,27\,-\,SMM     & $3.0\pm0.5$ & ($6.5\pm1.5$)E21 & 0.19 \\
B\,68\,-\,SMM        & $2.2\pm0.3$ & ($1.5\pm0.4$)E21 & 0.06 \\
CB\,244\,-\,SMM2 & $2.5\pm0.4$ & ($5.0\pm1.0$)E21 & 0.35 \\
\hline
\end{tabular}
\tablefoot{
\tablefoottext{a}{Approximate 1\,$\sigma$\ uncertainties derived from the $\chi^2$\ maps.} 
\tablefoottext{b}{With two free parameters and 23\,--\,40 radial points, the system is formally overdetermined 
                               such that $\chi^2_\mathrm{red,min}\ll1$.}
}
\end{table*}

Table\,\ref{t_overview_results_1} lists for all six cores the individual best-fit parameter values $s^{\prime}_\mathrm{ISRF}$\ and 
$N^{\prime}_H(r_\mathrm{sym})$, along with the minimal $\chi^2_\mathrm{red}$\ values from comparing the ray-tracing 
and radiative transfer results.
From the $\chi^2$\ maps, we also derive approximate mean relative 1\,$\sigma$\ uncertainties of 
\mbox{$\sigma(s^{\prime}_\mathrm{ISRF})\approx 15-20$\%} and \mbox{$\sigma(N^{\prime}_H(r_\mathrm{sym}))\approx 25$\%} for all sources.
Despite the qualitatively different effects on the equilibrium temperature distribution from varying 
$s_\mathrm{ISRF}$\ and $N_\mathrm{H}(r_\mathrm{sym})$\ (Figs.\,\ref {fig_b68_field} and \ref{fig_b68_Ak}), 
the two parameters remain partially correlated and the uncertainties are therefore not completely independent.
However, we find that solutions with $s_\mathrm{ISRF}$\ deviating by more than a factor of two from our best-fit value 
do no longer fit the data at all. 
This should be kept in mind when interpreting the absolute uncertainty values for the individual sources 
listed in Tab.\,\ref{t_overview_results_1}, 
which we derived from the mean relative uncertainties mentioned above.

The corresponding comparison between the temperature profiles from the two methods is shown in Fig.\,\ref{fig_bestmodels}.
The quality of the fits differs between the sources, which is reflected in both the $\chi^2_\mathrm{red,min}$\ values 
(Tab.\,\ref{t_overview_results_1}) and in the plots (Fig.\,\ref{fig_bestmodels}). 
The profiles for CB\,4 and B\,68 show excellent agreement.
For CB\,27 and CB\,244, the respective profile shapes differ slightly, though the predicted equilibrium 
temperatures are still within the uncertainty range of the ray-tracing results of $^{+2}_{-1}$\,K 
everywhere within $r_\mathrm{sym}$\ (Sect.\,\ref{ssec_disc_uncert}). 
For CB\,17 and CB\,26, this latter quality criterion could not be strictly met. 
In both cores we 'measure' (with the ray-tracing inversion) a smaller temperature gradient than what is predicted 
by the radiative transfer models for even the largest allowable values of $N_H(r_\mathrm{sym})$. 
In Sect.\,\ref{ssec_disc_uncert_sources} we discuss possible reasons for these discrepancies.

With the exception of CB\,4 ($s^{\prime}_\mathrm{ISRF}=1.1\pm 0.3$), 
the most distant source in our sample (Table\,\ref{t1}) and the only core with clearly subcritical 
central density (Fig.\,\ref{fig_massdens}), we find that the GALPROP ISRF 
generally results in temperatures that are significantly lower than what we derive from the data with the ray-tracing inversion.
Indeed, the other five sources were best fit with $\langle s^{\prime}_\mathrm{ISRF}\rangle=2.6\pm 0.4$.
This trend was also found by, for example,  \citet{shirley2005} and \citet{launhardt2013}.
However, we must caution that this finding only holds for the applied dust opacity model 
and dust models with other NIR--to--FIR opacity ratios may lead to other values of $s^{\prime}_\mathrm{ISRF}$\ 
(see discussion of uncertainties in Sect.\,\ref{ssec_disc_uncert_dust}). Therefore, we refrain here from drawing 
general conclusions on the total strength of the local ISRF. The lower value for CB\,4 may thus not necessarily suggest 
a weaker local ISRF, but may well be related to a different dust opacity law in this low-density cloud.

For most sources, we also had to increase the envelope extinction to the maximum value allowed by the observational constraints
(Table\,\ref{t_coreproperties}). However, since the estimated uncertainty of the best-fit $N_H(r_\mathrm{sym})$\ values is 
of the same order as the uncertainty of the observational constraints, we do not consider this trend significant.


\section{Discussion}     \label{sec_disc}

\subsection{Uncertainties}     \label{ssec_disc_uncert}

We estimate the total uncertainty of the dust temperature in the starless cores to be 
$\sigma_T\approx ^{+2}_{-1}$\,K at $T\approx 10$\,K along the entire radial profile of the sources.
The most significant contribution to the uncertainty of the central temperature was found to come from 
the symmetry assumptions made in the ray-tracing inversion method.
The most significant contribution to the uncertainty of the outer (envelope) temperature comes from the 
irregular geometry of the envelopes. 
The relative uncertainty of the central density is estimated to be $\sigma_{n,\mathrm{rel}}\approx35$\%,
with the most significant contributions coming from the inversion method and the dust opacities. 
However, this latter statement assumes that we consider only reasonably applicable dust opacity models 
and not the entire range of the still observationally poorly constrained dust opacity models that are in the literature. 
The relative uncertainty of the values for the total core mass is estimated to be $\sigma_{M,\mathrm{rel}}\approx30$\%
The formal uncertainty on the derived relative strength of the ISRF is estimated to be $\sigma_{s_\mathrm{ISRF}}\approx20$\% 
when considering the modified OH5a dust opacity model only. However, as mentioned above (Sect.\,\ref{ssec_res_comp}) 
and discussed further in Sect.\,\ref{ssec_disc_uncert_dust}, the actual uncertainty of the ISRF strength may be significantly larger 
when dust opacity models with different NIR--to--FIR opacity ratios are allowed.
In the following, we discuss separately the contributions to these uncertainties coming from the data and the ray-tracing inversion method,
the source properties, the dust opacity law,  and the assumed ISRF.


\subsubsection{Data and inversion method}     \label{ssec_disc_uncert_data}

The uncertainties introduced by the observing methods and data calibration and reduction procedures are analyzed 
and discussed in detail in \citet{launhardt2013}. Although the flux calibration of the {\it Herschel} data has been improved 
since then, the uncertainties are dominated by data reduction issues such as, for example, spatial filtering 
and background determination, such that the assessment in \citet{launhardt2013} is still considered valid. 
Hence, we estimate the data-related uncertainty of the dust temperatures inferred by the ray-tracing 
inversion to be $\sigma_T<\pm0.5$\,K at 10\,K.

The uncertainties of the temperature mapping and ray-tracing inversion method have already been assessed in 
three preceding papers \citep{nielbock2012,launhardt2013,lippok2013}. 
The uncertainties of the dust temperature estimation in the envelopes, where LoS temperature gradients are small, 
are dominated by the data-related uncertainties (see above). 
Towards the core centers, where flux densities are much higher, but also LoS temperature gradients,
the local uncertainty of the derived core temperature is dominated by the fine-tuning of the profile parameters 
in the ray-tracing inversion, which is related to the symmetry assumptions and the imperfect convergence between 
PoS and LoS profile parameters (Sect.\,\ref{ssec_mod_rti}). 
Based on the analysis in \citet{nielbock2012}, we estimate the inversion method-related uncertainty of the central 
core temperatures to be $\sigma_{T_0}\approx^{+2}_{-1}$\,K. 

Since the uncertainty of the LoS mean (optical depth-weighted) temperature is 
much smaller than that of the local central temperature \citep[see][]{launhardt2013}, 
the method-related uncertainty of the column density is also small. 
From Monte Carlo tests, we estimate it to be $\sigma_{N_0}\lesssim\pm10$\%.
Due to this balancing along the LoS, the method-related effect on the value of the central density 
is also smaller than one would infer from the uncertainty of the central temperature only. 
From Monte Carlo tests, we estimate it to be $\sigma_{n_0}\lesssim \pm20$\%.
However, the latter two statements only hold for the most nearby and fully spatially resolved cores
as we explain below.

We also tested the effects of beam smoothing by omitting the 500\,$\mu$m data and working at the 
350\,$\mu$m resolution of 25\arcsec, since the 500\,$\mu$m {\it Herschel} beam of 36\farcs4 \citep{aniano2011}
might not fully resolve the coldest parts of the core centers \citep[see also][]{nielbock2012,schmalzl2014}.
For example, for the two most distant sources in our sample (CB\,4 and CB\,17), the flat-density core diameters
($2\times r_0$, Tab.\,\ref{t_coreproperties}) are only $\approx$2.7 times larger than the beam.
We did indeed find an increase of the central density and peak column density by up to 50\% 
for these two sources when working with the smaller beam.
However, the resulting values of the central temperature were always within this uncertainty range and we 
did not find any systematic trend towards lower temperatures.
Since the temperature rises faster towards larger radii than the density decreases, that is, the radial temperature 
profiles are typically 30-40\% narrower than the density profiles, this effect does not arise from unresolving the cold, 
flat-density core. It can rather be attributed to smoothing the steep drop-off of the radial temperature profiles 
and thus effectively re-distributing the mass along the LoS in the radius range $\approx 5000 - 20000$\,au 
(see Fig.\,\ref{fig_dustprofiles}). 
The amplitude of these effects, which are related to the aforementioned fine-tuning and convergence 
of PoS and LoS profile parameters, thus depends on the angular size and morphology of the individual cores 
and cannot be uniquely quantified. 
This analysis also illustrates one short-coming of our method and the advantage of forward-modeling which would 
take full advantage of the higher angular resolution of the shorter-wavelength data.


\subsubsection{Source-specific properties}     \label{ssec_disc_uncert_sources}

Even in the most simple-structured star-forming regions like Bok globules, the envelopes of the starless cores 
are not strictly spherical \citep[e.g.,][]{myers1991,launhardt2013,lippok2013}. 
Furthermore, three out of six globules in our sample contain a nearby secondary core with an embedded protostar. 
In several cases, the starless cores are also not located in the center of the globules.
We partially eluded these problems by masking those regions that clearly deviate from the symmetric structure. 
We also set an outer radius, $r_\mathrm{sym}$, for 
the comparison of ray-tracing and radiative transfer results where the assumption of symmetry breaks down or where 
the signal in the continuum maps gets too weak to allow for reliable derivations of the structure. While this 
procedure is to some extent arbitrary, we verified that the results 
on $s^{\prime}_\mathrm{ISRF}$\ and $N^{\prime}_H(r_\mathrm{sym})$\
are nearly unaffected when varying $r_\mathrm{sym}$\ within reasonable ranges.
Nevertheless, this radially increasing deviation from spherical symmetry leads to an uncertainty 
also for the outer parts of the azimuthally averaged temperature profiles.
From the azimuthal scatter of the best-fit mid-plane temperature values at $r_\mathrm{sym}$, 
we estimate this uncertainty to be also $\sigma_{T_\mathrm{out}}\approx^{+2}_{-1}$\,K.

The most important specific properties of the individual globules are described in \citet{launhardt2010} and \citet{launhardt2013}
and summarized in Table\,\ref{t1}.
Here we only discuss those characteristics that are relevant for the interpretation of results from this paper.
CB\,4 and B\,68 are the two cores where the temperature profiles from the ray-tracing technique and the radiative transfer 
models had excellent agreement.  These two cores are also the most round ones. 
Additionally, these sources are both single cores\footnote{Ignoring the very low-mass secondary core at the 
tip of the south-western trunk in B\,68 \citep[see][]{alves2001a,nielbock2012}.}
without a nearby protostar, suggesting that they have the simplest structures in our sample.  

Conversely, CB\,27 and CB\,244\,-\,SMM2, both of which showed moderate agreement between the two modeled temperature profiles, 
are more complex.  CB\,27 is the most elliptical globule in our sample (see Fig.\,\ref{fig-tmaps}), 
which may explain the deviations between the 1-D radiative transfer models and the ray-tracing results.  
CB\,244\,-\,SMM2 has a nearby bright protostellar core (SMM1, 90\arcsec\ or 18,000\,au to the west), 
and our separation of the two cores may not have been perfect, such that the recovered dust emission toward 
the starless core may be enhanced by the presence of this protostar.  
Therefore, the ray-tracing inversion technique may not accurately portray the temperature profile of this source.
In addition, but to a smaller extent, anisotropic heating of the starless core by this protostar may also play a role.
However, we do not consider the local enhancement of the radiation field due to the protostar in SMM1 significant 
since it is located at a projected distance of 18,000\,au, has an estimated bolometric luminosity of only 
$\approx$1.8\,L$_{\odot}$\ \citep{launhardt2013}, and its outflow cones (through which most of the NIR/MIR luminosity escapes) 
are oriented perpendicular to the connecting line towards SMM2 \citep[see][]{yc1994}.

Finally, for CB\,17\,-\,SMM and CB\,26\,-\,SMM2, the ray-tracing technique and the radiative transfer models produced inconsistent temperature profiles.  
For CB\,17\,-\,SMM, a low-luminosity Class\,I young stellar object (YSO) is located \mbox{$\approx10$\arcsec}\ (2,500\,au) northwest 
from the center of the starless core \citep{schmalzl2014}. 
This YSO impairs the flux profile reconstruction for the starless core and leads to larger systematic 
deviations than for the other globules as this blend probably leads to an overestimation of the inner temperature of the starless 
core in the ray-tracing inversion. 
This YSO may also locally enhance the radiation field to which the core SMM is exposed. 
However, as in the case of CB\,244, we consider this effect less significant than the blending mentioned above since the YSO has a bolometric luminosity 
of only 0.12\,L$_{\odot}$\ \citep{launhardt2013} and is not embedded in the core SMM \citep{chen2012}.
In addition, the core of CB\,17\,-\,SMM is less resolved than most other globule cores, 
because of the combination of a relatively small physical size and a large distance (250\,pc), 
such that beam-smearing may actually not be negligible for this core. 

CB\,26 is a globule with multiple cores and a relatively complex visual appearance \citep[][Fig.\,C.4]{launhardt2013}. 
Moreover, C$^{18}$O(2\,--\,1) spectra of the SMM2 core show multiple velocity components \citep[][Fig.\,3]{lippok2013} 
which suggest that this core may actually be a pole-on view of a longer filament or the projection of two or more diffuse cores. 
If CB\,26-SMM2 is indeed a by-chance projection of two or more diffuse cores in a filament seen pole-on, 
both the ray-tracing inversion and the \mbox{1-D} radiative transfer would fail to 
reproduce its physical structure and internal temperature distribution. 
For this reason, we do not further consider CB\,26-SMM2 for the following discussion 
of the effects of the dust opacity law and the ISRF.


\subsubsection{Dust opacities}     \label{ssec_disc_uncert_dust}

The derivation of a temperature from the SED of the thermal dust emission is sensitive to the spectral slope $\beta$\
of the dust opacity law at FIR through mm wavelengths. 
However, most previous observations of cold and dense molecular cloud cores lack robust detections over this wide 
wavelength range, resulting in poorly constrained estimates of $\beta$.  
Thus, many studies instead adopt appropriate modeled dust opacity laws, 
most of which have slopes in the range \mbox{$\beta\approx1.9\pm0.1$}\ \citep[e.g.][]{oh94,weingartner2001,ormel2011}.
Within this range, the dust temperatures inferred from the SEDs would vary by less than $\pm0.5$\,K \citep[see][]{launhardt2013}.
However, the combination of these slight temperature differences with the much larger differences in the absolute values of the 
FIR\,--\,mm dust opacities between the various dust models results in an uncertainty of the density, column density, and mass 
of about a factor of three. This would in turn affect the shielding and energy balance of the cores and thus the predicted equilibrium 
temperature distributions and $s_\mathrm{ISRF}$\ values (see also discussion below).
Although the absolute scaling value for the FIR\,--\,mm opacity is unknown, one can generally exclude models with unprocessed
ISM-type dust grains or models with extremely coagulated grains that also have very thick ice mantles.  
Such extreme cases are unlikely to represent the conditions in molecular cloud cores.

Ideally, we would use different dust opacity laws for the core centers and envelopes, as both regions differ in temperature and density
and hence in the expected degree of grain processing.
While the use of two different dust opacity laws would have little effect on the observed dust temperature profiles, 
there would be a more significant effect on the density profiles 
(and hence on the equilibrium temperature distribution and ISRF strength inferred from radiative transfer models).
We tested such a two-layered dust model on B\,68, using OH5a opacities at $n_\mathrm{H}>10^4$\,cm$^{-3}$\ 
\mbox{($\cong r>1.5\,10^4$\,au)} and OH1 opacities outside.
Compared to our single-opacity results with OH5a opacities alone (see Table\,\ref{t_coreproperties}), 
we found that the two-layered model gave consistent results for $T_0$,
a 0.8\,K increase for $T_\mathrm{out}$, a 25\% increase of $n_0$, a 50\% increase of $N_0$ , and a lower value of $\eta$\ (4 instead of 5).
This may partially explain our relatively large values of $\eta$\ in Table\,\ref{t_coreproperties}.
Regarding the best-fit values of $\eta$, it should also be mentioned that there is a relatively strong degeneracy between 
$r_0$\ and $\eta$, such that the $\eta$\ values listed in Table\,\ref{t_coreproperties} should not be over-interpreted.
For the two-layered opacity model of B\,68, we could, for example, still obtain reasonable fits to the data by forcing 
$r_0$\ to 20\% below the formal best-fit value, which would decrease the value of $\eta$\ further from four to three.
Nevertheless, we lack sufficient constraints to use such a multi-layered dust model, and we caution that our comparison 
for B\,68 is not robust quantitatively. It only qualitatively reveals the trend of the systematic uncertainties introduced 
by using a single-layer dust opacity model for such cloud cores.

\begin{figure}[htb]
   \centering
   \includegraphics[width=0.45\textwidth]{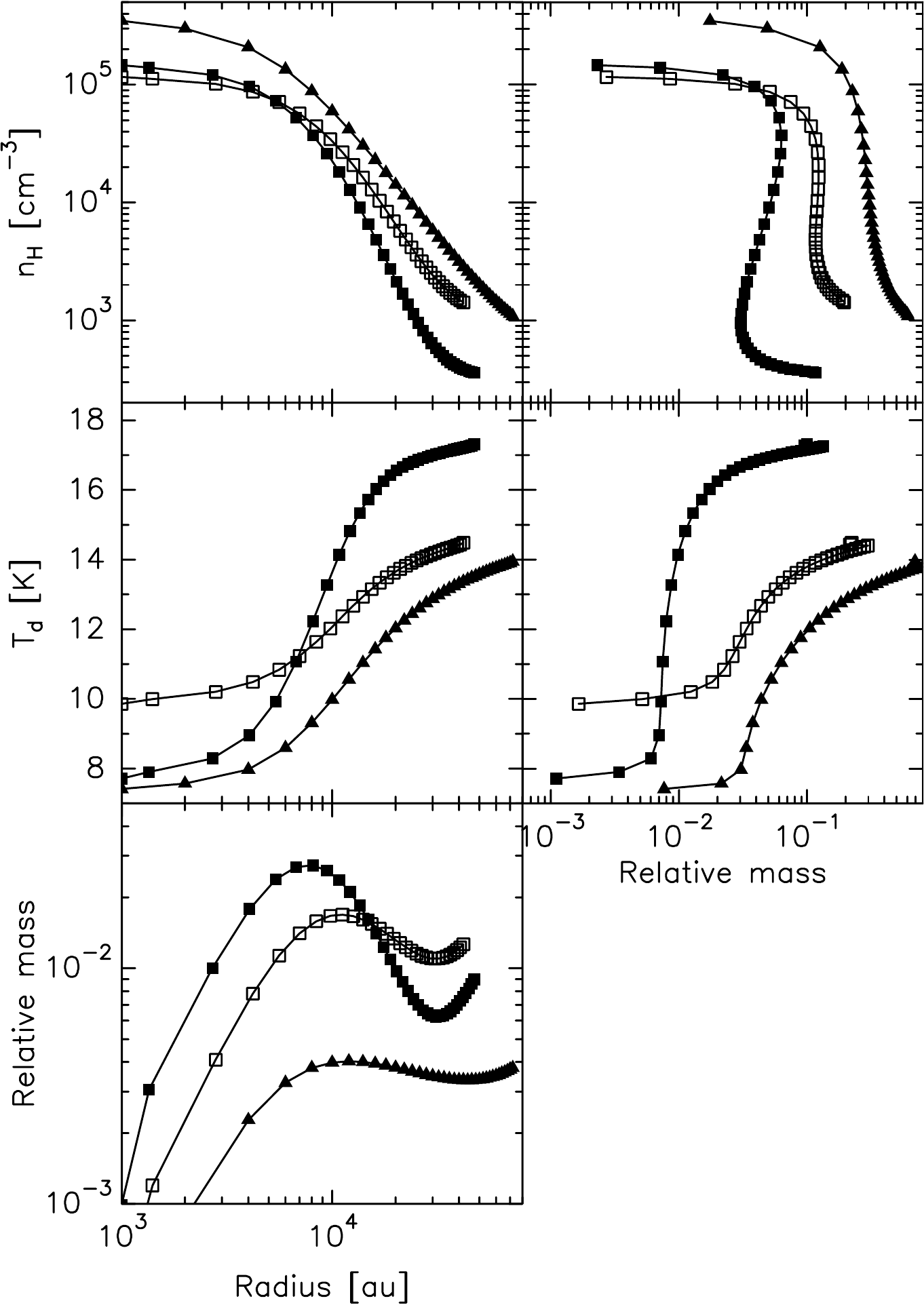}
   \caption{Radial profiles of volume density, dust temperature, and relative mass per 
    $\Delta R=1000$\,au shell (left from top to bottom), as well as relative mass per logarithmic density interval 
    and per temperature interval $\Delta T=1$\,K (right) for B\,68 (filled squares), CB\,27 (open squares) 
    and CB\,244\,-\,SMM2 (filled triangles).}
   \label{fig_massprofile}
\end{figure}

This well-known problem has been discussed recently by, for example, \citet{pagani2015}.
Their analysis, like ours above, shows that the lack of good constraints on the dust opacity law and its possible change 
along the LoS can lead to significant uncertainties in the column densities and masses derived from dust emission data
alone, in particular when very cold ($T_\mathrm{in}<10$\,K) cores are involved. 
However, our more sophisticated treatment of the temperature structure of the starless cores as compared to \citet{pagani2015} 
leads to smaller uncertainties than derived by these authors. 
Furthermore, we can show that the larger uncertainty on the values of the 
dust opacity and temperature in the central core region with high density has 
very little effect of the estimate of the total core mass.
The radial profile and mass distribution diagrams for B\,68, CB\,27, and CB\,244\,-\,SMM2 (Fig.\,\ref{fig_massprofile})
clearly demonstrate that the cold and dense innermost regions of this core 
actually contribute very little to the total mass. Most of the core mass 
comes from a shell with $\approx 5000-15000$\,au radius, where the dust temperature is 
\mbox{$\approx 2-5$\,K} higher than at the core centers and the densities are a few \mbox{$10^4$\,cm$^{-3}$}. 
If expressed in temperature intervals (e.g., 1\,K), most of the mass actually 
sits in the outer low-density envelope where the dust temperature is $\ge13-15$\,K. 
This shows that, despite of the decreased sensitivity and increased uncertainty for the temperature 
and column density of the coldest dust, there is in practice no 'hidden mass' problem in such cores.
Nevertheless, the degeneracy between dust opacity and temperature cannot be solved with dust emission data alone 
and we can only derive an estimate of the related uncertainties in the derived column densities and masses.

Furthermore, the NIR--to--FIR opacity ratio is another key parameter of the dust opacity law.
Although this ratio does not affect the derivation of a temperature from the thermal SEDs, 
it greatly affects the shielding of the cloud from the ISRF and the subsequent energy balance, 
which in turn affects the equilibrium temperature distribution predicted by self-consistent radiative transfer models
and thus the ISRF strength deduced.
This NIR--to--FIR opacity ratio is very uncertain, and the aforementioned opacity models have large scatter; 
they vary within a factor of four. 
Since different models have different trends when going from ISM-type to more processed dust,
we cannot easily predict in which direction the radiative transfer temperature profiles would change and how 
the comparison with the ray-tracing results would be affected. 
Consequently, we cannot exclude currently that our conclusions on the value of $s^{\prime}_\mathrm{ISRF}$\ 
(Sect.\,\ref{ssec_res_comp}), which we derive from comparing the radiative transfer temperature profiles with the 
ray-tracing results, might be significantly affected by this uncertainty in the NIR--to--FIR opacity ratio of the applicable dust model.
Thus, we need better observational constraints on the NIR--to--FIR opacity ratio, a subject which 
we will analyze and discuss in more detail in a forthcoming paper.

Related to this and continuing in the same theme is the uncertainty in the applicable albedos (or scattering efficiencies) mentioned in Sect.\,\ref{ssec_mod_dust}.
The difference between the applied WD3.1 albedos and the possible alternative WD5.5B \citep{weingartner2001} would be 
an about 60\% increased extinction efficiency at the peak of the albedo spectrum at \mbox{$\lambda\approx1\,\mu$m}.
Although this is much smaller than the aforementioned general uncertainty in the NIR--to--FIR opacity ratio, 
the application of WD5.5B albedos instead of those from WD3.1 would result in systematically slightly increased shielding of the 
cold core centers in the radiative transfer modeling and thus imply the derivation of even higher values of $s_\mathrm{ISRF}$\ 
(see also discussion in Sect.\ref{ssec_disc_uncert_isrf}.)

Finally, laboratory experiments indicate that the dust opacities in the FIR and mm-range have a more complicated behavior 
than assumed by current dust models. \citet{boudet2005} and \citet{coupeaud2011} showed that the spectral index 
$\beta$\ in the FIR and mm-range depends on temperature and on wavelength. Both studies found an increase of 
$\beta$\ above two at low temperatures and at wavelengths above a few hundred $\mu$m, depending on the 
composition of the grains. These findings can be explained with a quantum mechanical model of amorphous 
grains \citep{paradis2011}. Since robust observational constraints on such a (predicted) behavior are still lacking, 
we do not take into account this effect, but only mention here that it could contribute additional uncertainties.
Although the changes would probably be small compared to other uncertainties, this effect would slightly lower 
the central core temperature inferred from the SED (by at most a few tenths of K),
slightly increase the local core density, and also slightly affect the energy balance in the self-consistent 
radiative transfer modeling.


\subsubsection{Interstellar radiation field}     \label{ssec_disc_uncert_isrf}

Not only the cloud structure is asymmetric, the IRSF can also be. It could be enhanced towards the 
direction of luminous stars, star-forming regions, or the galactic plane. On the other hand, large molecular clouds 
located between the galactic plane and the globules can shield the globules from the general ISRF. In fact, it was found in 
\citet{launhardt2013} that most globules have azimuthally varying outer temperatures of their envelopes
with one side often being warmer than the other. 
However, \citet{launhardt2013} did in general not find a connection between directions of increased temperatures and 
galactic heating sources. In our modeling approach, we should not introduce significant systematic errors 
into the analysis by assuming an isotropic ISRF, because local variations of the ISRF should be coupled to local 
variations of the dust temperatures and thus an azimuthally averaged temperature profile should to first order 
correspond to the azimuthally averaged ISRF.

We also estimated the dependence of the results on the spectral shape of the ISRF by making the same analysis as 
described in Sec.\,\ref{ssec_res_comp} for radiative transfer models that assume the ISRF model by 
\citet[][see also Fig.\,\ref{fig_ISRF}]{black1994}. We find that $\langle s_\mathrm{ISRF}\rangle$\ is 
by a factor $1.3 - 1.5$\ higher for these models than for models assuming the GALPROP ISRF, which 
roughly corresponds to the difference in the flux at wavelengths shortward of 8\,$\mu$m 
(Sect.\,\ref{ssec_mod_srt} and Fig.\,\ref{fig_ISRF}). 
The relative shape of the temperature profile (in--out contrast) is nearly unaffected. 
Although the comparison with the \citet{black1994} ISRF comprises a very simple and limited test 
of the effects from varying the spectral shape of the ISRF, we conclude that the uncertainty in our knowledge of the exact spectral 
shape of the ISRF is negligible compared to the uncertainty of its total flux at UV to IR wavelengths.
Furthermore, as discussed in the previous section, we do not consider our results on the absolute strength of the ISRF robust 
since the degeneracy with the observationally only poorly constrained NIR--to--FIR dust opacity ratio cannot be resolved 
based on the available data.


\subsection{Comparison with previous studies}     \label{ssec_disc_prev}

\begin{table*}[htb]
\caption{\label{t_comparison_previous}
Comparison of $N_\mathrm{H}$\ peak column densities with previous dust emission studies.}
\centering
\begin{tabular}{lccccccl}
\hline\hline
Reference         & 
CB\,4                  &                
CB\,17                & 
CB\,26                 & 
CB\,27                  & 
B\,68                     & 
CB\,244                & 
Remarks                   \\
                             &
                             &
SMM                     &
SMM2                   &
                             &
                             &
SMM2                   &
                              \\
\hline
lau97a      & $\ldots$ &  4.3e22 & $\ldots$ & $\ldots$ & $\ldots$ & $\ldots$ &  1.3\,mm on-off, T$\equiv$20K\tablefootmark{a}, $\kappa_\mathrm{0}\tablefootmark{e}=0.008$\,cm$^2$\,g$^{-1}$ \\
lau97b      & $\ldots$ &  2.9e22 & $\ldots$ & $\ldots$ & $\ldots$ & $\ldots$ & 0.45\,--\,1.3\,mm on-off, T$\equiv$20K\tablefootmark{a}, $\kappa_\mathrm{0}=0.008$\,cm$^2$\,g$^{-1}$ \\
lau10        & $\ldots$ &  2.0e23 & $\ldots$ & $\ldots$ & $\ldots$ & 3.0e23  &  0.45\,--\,1.2\,mm maps, T$\equiv$10K\tablefootmark{a}, $\kappa_\mathrm{0}=0.005$\,cm$^2$\,g$^{-1}$  \\
stutz10     & $\ldots$ & $\ldots$ & $\ldots$ & $\ldots$ & $\ldots$ & 8.0e22  & 0.1\,--\,1.2\,mm maps, mBB fit\tablefootmark{b}, $\kappa_\mathrm{0}=0.005$\,cm$^2$\,g$^{-1}$  \\
lau13        & 7.5e21  & 2.5e22   & 1.4e22  & 1.9e22   & 2.5e22  & 4.6e22   & 0.1\,--\,1.2\,mm maps, mBB fit\tablefootmark{b}, $\kappa_\mathrm{0}=0.009$\,cm$^2$\,g$^{-1}$  \\
nielb12     & $\ldots$ & $\ldots$ & $\ldots$ & $\ldots$ & 4.3e22 & $\ldots$ & 0.1\,--\,1.2\,mm maps, RTI\tablefootmark{c}, $\kappa_\mathrm{0}=0.007$\,cm$^2$\,g$^{-1}$ \\
lipp13        & 9.2e21\tablefootmark{f}  & 3.0e22\tablefootmark{f}   & 2.0e22  & 2.7e22   & 5.5e22 & 9.1e22   & 0.1\,--\,1.2\,mm maps, RTI\tablefootmark{c}, $\kappa_\mathrm{0}=0.007$\,cm$^2$\,g$^{-1}$ \\
roy14         & $\ldots$ & $\ldots$ & $\ldots$ & $\ldots$ & 2.6e22 & $\ldots$ & 0.16\,--\,0.5\,mm maps, AI\tablefootmark{d}, $\kappa_\mathrm{0}=0.009$\,cm$^2$\,g$^{-1}$ \\
schma14   & $\ldots$ & 4.3e22  & $\ldots$ & $\ldots$ & $\ldots$ & $\ldots$ & 0.1\,--\,1.2\,mm maps, RTI\tablefootmark{c}, $\kappa_\mathrm{0}=0.007$\,cm$^2$\,g$^{-1}$ \\
This work  & 1.0e22\tablefootmark{f}   & 3.0e22\tablefootmark{f}  & 1.7e22   & 2.6e22  & 2.8e22   & 9.2e22  & 0.1\,--\,1.2\,mm maps, RTI\tablefootmark{c}, $\kappa_\mathrm{0}=0.007$\,cm$^2$\,g$^{-1}$ \\
\hline
\end{tabular}
\tablebib{lau97a:~\citet{launhardt1997AA}; 
lau97b:~\citet{launhardt1997MNRAS}; 
lau10:~\citet{launhardt2010}; 
stutz10:~\citet{stutz2010}; 
lau13:~\citet{launhardt2013}; 
nielb12:~\citet{nielbock2012};
lipp13:~\citet{lippok2013}; 
roy14:~\citet{roy2014}; 
schma14:~\citet{schmalzl2014}.
}
\tablefoot{
\tablefoottext{a}{$T_\mathrm{d}$\ fixed due to lack of observational constraints.} 
\tablefoottext{b}{Modified blackbody fit to LoS-averaged $T_\mathrm{d}$.}
\tablefoottext{c}{Ray-tracing inversion, accounting for LoS $T_\mathrm{d}$\ variation.}
\tablefoottext{d}{Abel inversion, accounting for LoS $T_\mathrm{d}$\ variation.}
\tablefoottext{e}{For direct comparability, we list here the total (gas\,+\,dust) opacity at $\lambda=1.0$\,mm,  $\kappa_\mathrm{0}$, of the ISM model used in the respective paper.}
\tablefoottext{f}{Peak column density probably underestimated by up to 30\% due to beam-smoothing; see discussion in Sect.\,\ref{ssec_disc_uncert_data}.}
}
\end{table*}

In Table\,\ref{t_comparison_previous}, we compiled a comparison of $N_\mathrm{H}$\ peak column densities 
(i.e., $N_\mathrm{H_2}$\ converted where applicable) derived from this work with earlier studies of the dust emission from the respective sources.
We note that, in addition to the differences in wavelength coverage, data quality, and treatment of the dust temperature, the studies also differ in angular resolution 
(12\arcsec\,--\,36\arcsec) and dust opacity law used (see last column in Tab.\,\ref{t_comparison_previous}).
Nevertheless, the comparison shows that,
with the few exceptions discussed below, the results of the different studies are consistent with each other if one considers 
the methods involved, that is, simple single-temperature SED fits result in somewhat lower column densities than the ray-tracing inversion 
and assumptions of a too low fixed temperature results in higher column densities. 
The $\approx40$\% higher column density derived by \citet{schmalzl2014} for CB\,17 
can be attributed entirely to the different angular resolutions that were used in these two studies
(25\arcsec\ vs. 36\arcsec; see also discussion in Sect.\,\ref{ssec_disc_uncert_sources}).
The higher column densities derived for B\,68 by \citet{nielbock2012} and \citet{lippok2013} 
are not related to a smaller beam size, but can only be attributed to the problems of fine-tuning 
the profile parameters when the steep radial temperature gradient is not fully resolved 
(see discussion in Sect.\,\ref{ssec_disc_uncert_sources})
and the improved PoS\,--\,LoS convergence scheme for the ray-tracing algorithm mentioned in 
Sect.\,\ref{ssec_res_rti} and used for this paper.

To our knowledge, B\,68 is the only core in our sample with a previous characterization of its dust temperature structure by other authors.
\citet{bianchi2003} compared SCUBA~850\,$\mu$m and SIMBA~1.2\,mm maps 
of this globule to the NIR extinction map of \citet{alves2001b} and calculated the dust temperature distribution 
with a model by \citet{goncalves2004}. The correlation of the dust emission to the NIR extinction showed a 
flattening toward large $A_\mathrm{K}$\ and they showed that models assuming a positive temperature gradient can 
describe the data better than models assuming an isothermal cloud, although their data did not constrain 
the temperature of B\,68. 
\citet{bergin2006} also calculated the dust temperature distribution of B\,68 using the radiative transfer 
code of \citet{zucconi2001} and the density structure that was derived by \citet{alves2001b} from NIR extinction measurements. 
They deduced a central dust temperature of about 8\,K, which agrees well with our analysis.
Recently, \citet{roy2014} used an inverse-Abel transformation based technique to infer the dust temperature and density 
structure of B\,68 from {\it Herschel} data. Apart from the treatment of the outer halo (Sect.\,\ref{ssec_mod_rti} 
and Eq.\,\ref{eq_plummer}), and the use $n_\mathrm{H_2}$\ instead of $n_\mathrm{H}$, their results agree well with ours 
from this paper.
In \citet{nielbock2012}, we also used radiative transfer modeling to test the hypothesis that the increased FIR emission 
on the side facing the galactic plane \citep[see Fig.\,C.8 in][]{launhardt2013} could be the result of a stronger ISRF 
coming from this direction. While the models supported this assumption qualitatively, they could not constrain the degree of 
anisotropy in the ISRF.

Another study that investigated the density and temperature structures of 20 dense cores in the L\,1495 cloud in Taurus 
based on \textit{Herschel} data was recently presented by \citet{marsh2014}. Although there is no overlap between their 
sources and ours, their COREFIT algorithm differs from our ray-tracing inversion in several aspects 
(e.g., forward-modeling vs. inversion starting from data, assumption of spherical geometry vs. allowing for moderate deviations 
from even elliptical geometry, etc.), and they use a different model for the ISRF from us, these authors derive conclusions that 
are very similar to ours. 
They estimate central core dust temperatures in the range $6-12$\,K (our range spans $7.5-11.9$\,K) that are also negatively 
correlated with peak column density, suggesting external heating by the ISRF and dust shielding of the central cores. 
These authors also encounter the problem that radiative transfer models require un up-scaling of the ISRF 
with respect to the COREFIT results to explain their central core temperatures. 
However, due to the different modeling approaches and ISRF models involved, a quantitative comparison with our results 
is difficult. As in our paper, they don't find a conclusive explanation for this behavior, but can only suggest that further study 
is necessary.

Last but not least, we want to mention that other authors have also attempted to derive gas kinetic temperature distributions 
of similar starless cores. For example, \citet{pagani2007} employed 1-D non-LTE molecular line radiative transfer modeling to infer 
the radial profile of the kinetic gas temperature in the L\,183 starless core from observations of N$_2$H$^+$\ and N$_2$D$^+$.
They find that that the gas in the core center is very cold ($7\pm1$\,K) and thermalized with the dust and increases outward 
up to about 12\,K at $1.3\times10^4$\,au. While both the trend of outward increasing temperature as well as the absolute 
values of the temperature compare well to our results on the dust temperature in our cores, a direct comparison is difficult for 
the following reasons. Due to the abundance and excitation profiles of the respective molecules \citep[see also][]{lippok2013}, 
the sensitivity at larger radii, where we still can measure dust temperatures, is lost. Furthermore, at larger radii and lower densities, 
the dust and gas are no longer thermally coupled and the gas temperature is expected to rise significantly above the dust temperature.


\section{Summary and conclusions}   \label{sec_sum}

With the \textit{Herschel} space observatory, it has become possible to spatially resolve nearby starless cores 
in the FIR and to spectrally sample the peak of the SED of the thermal dust radiation with high sensitivity for the first time. 
Such observations allow us to break the density-temperature degeneracy in the interpretation of their 
dust continuum emission and to derive tight constraints on the dust temperature and density distribution of the cores.
For this purpose, we have developed a ray-tracing inversion technique with which we 
can reconstruct the 3-D temperature and density structure of starless cores directly from the data with a minimum 
of assumptions and without restriction to a specific physical model \citep{nielbock2012,lippok2013,schmalzl2014}.

In this paper, we use Herschel observations of six starless cores from the EPoS sample to compare the observed temperature 
profiles from our ray-tracing inversion technique to the equilibrium temperature profiles from self-consistent 1-D radiative transfer models.
The only free parameters of the radiative transfer models are the relative strength of the ISRF (which we scale freely) 
and the selective extinction of the ISRF by a tenuous envelope (which we vary within the observational constraints). 
We restrict our analysis to one specific dust opacity model (OH5a), which we 
consider the most physically meaningful dust model for the core interiors, and only discuss how the application of other 
dust models would affect the results.

We derive physical outer radii of the six globules within the range $(6.5\pm2.5)\times10^4$\,au,
total core masses within a fixed radius of $5\times10^4$\,au in the range $2.6-14$\,M$_{\odot}$, 
central volume H number densities in the range $(2.5-32)\times10^4$\,cm$^{-3}$, 
peak column densities in the range $(1.0-9.2)\times10^{22}$\,cm$^{-2}$, 
and density profiles that are characterized by a flat-density core with radii in the range 
$(6-17)\times10^3$\,au, power-law fall-offs that are typically somewhat steeper than 
BES, and extended tenuous outer halos with typical dsnities of a few hundred H\,cm$^{-3}$.
We show that most of the core mass is neither located in the cold core centers, nor in the 
extended tenuous envelopes, but within a shell with $\approx 5000-15000$\,au radius, 
where the dust temperature is \mbox{$\approx 2-5$\,K} higher than at the core centers and the 
densities are a few \mbox{$10^4$\,cm$^{-3}$}. 
All starless cores are found to be significantly colder inside than outside.
Central core temperatures are in the range $7.5-11.9$\,K and show a strong negative correlation with  
peak column density. Outer envelope temperatures are in the range $13.5-19$\,K and 
core\,--\,envelope temperature differences are in the range $2.4-9.6$\,K. 
Altogether this suggests that the thermal structure of the cores 
is dominated by external heating from the ISRF and shielding by dusty envelopes.

We find that for four of the six cores, the dust temperature distributions inferred directly from the dust emission data with 
the ray-tracing inversion method can be reproduced well with self-consistent radiative transfer models. 
The best agreement is achieved 
for relatively round sources without nearby (blending) protostars, for which both the ray-tracing inversion
and 1-D radiative transfer work best (CB\,4 and B\,68). 
Slight discrepancies in the detailed temperature profile shapes, albeit within the uncertainty ranges, 
are encountered for sources that are very elliptical (CB\,27) or have nearby protostellar cores that partially overlap (in projection) 
with the starless core (CB\,244). In the first case, the 1-D approximation in the radiative transfer model was not optimal and 
it remains to be shown if the temperature structure inferred with the ray-tracing inversion can be reproduced 
better with 3-D radiative transfer.
In the second case, the blending by a nearby protostellar core and imperfect masking leads to larger uncertainties in the 
derived flux density profiles.

For two cores, the discrepancies between the temperature profiles inferred with the ray-tracing inversion
and 1-D radiative transfer exceed the formal uncertainties. 
Of these, CB\,17 was likely affected by strong blending by a nearby protostar, resulting in large uncertainties in the 
reconstructed flux density profiles of the starless core. 
For the other core, CB\,26, we found that this object may be a super-projection of three or more filamentary cores, 
such that neither 1-D radiative transfer is applicable, nor does the ray-tracing inversion result in a reliable temperature profile. 

We also confirm preliminary earlier results from various other studies which found that the usually adopted canonical value of the total strength 
of the ISRF in the solar neighbourhood is too low
when an OH5-type dust model is invoked. This becomes evident by radiative transfer models providing 
core temperatures that are lower than observed if the ISRF is not increased.
However, since the dust opacity model parameter that most strongly affects the conclusion on \mbox{$s^{\prime}_\mathrm{ISRF}$},
the NIR--to--FIR dust opacity ratio, is poorly constrained observationally, we cannot resolve this degeneracy and are unable to 
draw robust conclusions on the actual strength of the local ISRF.

In summary, we conclude that for cores with not too complex morphology and no blending by additional nearby sources, 
the ray-tracing inversion technique infers temperature and density profiles that can be 
well-reproduced with self-consistent radiative transfer models. Moreover, the ray-tracing inversion technique can naturally 
account for moderate deviations from spherical symmetry. 
Hence the method can also be applied to other cores, provided their geometry is not too complex and 
the data are of similar quality and cover a similar wavelength range to those in this study.
%


\begin{acknowledgements}
We thank A. Strong and T. Porter for advice regarding the details of the GALPROP ISRF model used in this paper. 
We also thank A.\,M. Stutz, Y.\,L. Shirley, J. Steinacker, and E. Keto for helpful discussions and comments on the paper draft. 
The work of J.K. and S.E.R. was supported by the Deutsche Forschungsgemeinschaft priority program~1573 
("Physics of the Interstellar Medium"). H.L. and M.N. were funded by the Deutsches Zentrum f\"ur Luft- und Raumfahrt~(DLR).
PACS has been developed by a consortium of institutes 
led by MPE (Germany), including UVIE (Austria); KU Leuven, CSL, IMEC (Belgium); CEA, LAM (France); MPIA (Germany); 
INAF-IFSI/OAA/OAP/OAT, LENS, SISSA (Italy); IAC (Spain). This development has been supported by the funding agencies 
BMVIT (Austria), ESA-PRODEX (Belgium), CEA/CNES (France), DLR (Germany), ASI/INAF (Italy), and CICYT/MCYT (Spain).
\end{acknowledgements}

\bibliographystyle{aa}
\bibliography{dustbib}


\end{document}